\pdfoutput=1
\documentclass[twoside,brazil,english,5p,sort&compress,hidelinks]{elsarticle}
\usepackage[T1]{fontenc}
\usepackage[latin9]{inputenc}
\usepackage{color}
\usepackage{babel}
\usepackage{booktabs}
\usepackage{calc}
\usepackage{amsmath}
\usepackage{graphicx}
\usepackage{subscript}
\usepackage[unicode=true,
 bookmarks=true,bookmarksnumbered=true,bookmarksopen=false,
 breaklinks=true,pdfborder={0 0 0},pdfborderstyle={},backref=false,colorlinks=true]
 {hyperref}
\hypersetup{pdftitle={ViscNet: neural network for predicting the fragility index and the temperature-dependency of viscosity},
 pdfauthor={Daniel R. Cassar},
 pdfkeywords={viscosity, neural network, machine learning, property prediction, feature extraction}}

\makeatletter

\newcommand*\LyXZeroWidthSpace{\hspace{0pt}}
\providecommand{\tabularnewline}{\\}

\journal{Acta Materialia (\href{https://doi.org/10.1016/j.actamat.2020.116602}{doi:10.1016/j.actamat.2020.116602})}

\usepackage{lineno}

\usepackage{url}

\usepackage[bracket-numbers = false,
  table-align-uncertainty = false, table-align-exponent=true]{siunitx}
\usepackage[version=3]{mhchem}

\usepackage{hyphenat}

\newcommand{\apdx}{Appendix}

\usepackage{fancyhdr}
\usepackage{lastpage}
\@ifundefined{showcaptionsetup}{}{%
 \PassOptionsToPackage{caption=false}{subfig}}
\usepackage{subfig}
\makeatother

\begin{document}

\begin{frontmatter}{}

\title{ViscNet: Neural network for predicting the fragility index and the
temperature-dependency of viscosity}

\author{Daniel R.\ Cassar}

\ead{contact@danielcassar.com.br}

\cortext[cor1]{Corresponding author}

\address{Department of Materials Engineering, Federal University of São Carlos,
São Carlos, Brazil}
\begin{abstract}
Viscosity is one of the most important properties of disordered matter.
The tem\-per\-a\-ture-dependence of viscosity is used to adjust
process variables for glass-making, from melting to annealing. The
aim of this work was to develop a physics-informed machine learning
model capable of predicting the tem\-per\-a\-ture-dependence of
the viscosity of oxide liquids, inspired by the recent Neural Network
(NN) reported by Tandia and co-authors. Instead of predicting the
viscosity itself, the NN predicts the parameters of the MYEGA viscosity
equation: the liquid\textquoteright s fragility index, the glass transition
temperature, and the asymptotic viscosity. With these parameters,
viscosity can be computed at any temperature of interest, with the
advantage of good extrapolation capabilities inherent to the MYEGA
equation. The viscosity dataset was collected from the SciGlass database;
only oxide liquids with enough data points in the \textquotedblleft high\textquotedblright{}
and \textquotedblleft low\textquotedblright{} viscosity regions were
selected, resulting in a final dataset with \num{17584} data points
containing 847 different liquids. About 600 features were engineered
from the liquids\textquoteright{} chemical composition and 35 of these
features were selected using a feature selection protocol. The hyperparameter
(HP) tuning of the NN was performed in a set of experiments using
both random search and Bayesian strategies, where a total of 700 HP
sets were tested. The most successful HP sets were further tested
using 10-fold cross-validation, and the one with the lowest average
validation loss was selected as the best set. The final trained NN
was tested with a test dataset of 85 liquids with different compositions
than those used for training and validating the NN. The coefficient
of determination ($R^{2}$) for the test dataset\textquoteright s
prediction was 0.97. This work introduces three advantages: the model
can predict viscosity as well as the liquids\textquoteright{} glass
transition temperature and fragility index; the model is designed
and trained with a focus on extrapolation; finally, the model is available
as free and open-source software licensed under the GPL3.
\end{abstract}

\end{frontmatter}{}

\noindent\fbox{\begin{minipage}[t]{1\columnwidth - 2\fboxsep - 2\fboxrule}%
\begin{center}
Publisher version available at \url{https://doi.org/10.1016/j.actamat.2020.116602}.
\par\end{center}
\begin{center}
Please cite this article as: Cassar, D.R. (2021). ViscNet: Neural
network for predicting the fragility index and the temperature-dependency
of viscosity. Acta Materialia 206, 116602. doi:10.1016/j.actamat.2020.116602
\par\end{center}%
\end{minipage}}

\section{Introduction}

Viscosity is one of the most important properties of disordered matter.
In the context of oxide glass-forming liquids, the tem\-per\-a\-ture-dependence
of viscosity is used to adjust process variables for glass making,
including conformation and annealing \citep{fotheringham2019viscosity};
it can also be used as a proxy for the diffusion coefficient for kinetic
processes such as crystal nucleation and crystal growth \citep{nascimento2010does,nascimento2011dynamic,cassar2017elemental,cassar2018diffusion}.
A new parameter of glass-forming ability was recently proposed based
on the viscosity at the liquidus temperature \citep{jiusti2020viscosity}.

Reliable predictive models are desired in practically all materials
science and engineering \citep{liu2014perspective}, including glass
science and technology \citep{mauro2018decoding}. These predictive
models are expected to increase the speed and reduce the cost of developing
new materials \citep{cassar2020designing}. This desire has increased
the interest in the interface between machine learning and oxide glass
science, as seen in a recent surge of publications on this topic \citep{mauro2016accelerating,guire2019datadriven,liu2019machine,tandia2019machine}.
In this context, the most used machine learning technique by far is
neural networks (NN) \citep{dreyfus2003machine,brauer2007solubility,bosak2016artificial,mauro2016accelerating,anoopkrishnan2018predicting,cassar2018predicting,ruusunen2018deep,bishnoi2019predicting,tandia2019machine,yang2019predicting,alcobaca2020explainable,deng2020machine,han2020machine,lillington2020predicting,onbasli2020mechanical,ravinder2020deep},
which are particularly good at finding patterns and modeling non-linear
dependencies between a set of features (input) and targets (output).
The usual approach found in the literature is to use a feedforward
NN as a universal regressor model to predict glass properties. This
approach is often referred to as a black-box, given the difficulty
of interpreting the internal rules of the model. 

Recently, Tandia et al.~\citep{tandia2019machine} developed a gray-box
approach to predict viscosity: they embedded a physical model in the
machine learning pipeline, which also contains a neural network. Compared
with the black-box approach, the gray-box approach improved the prediction
of viscosity by changing the purpose of the NN from a predictor of
viscosity to a predictor of the \emph{parameters} of a viscosity model,
the MYEGA viscosity model, shown in Eq.~\eqref{eq:MYEGA}. In the
MYEGA equation, $\eta$ is the viscosity, $T$ is the absolute temperature,
$\eta_{\infty}\equiv\lim_{T\rightarrow\infty}\eta\left(T\right)$
is the asymptotic viscosity, $m$ is the liquid's fragility index
(as defined by Angell \citep{angell1985strong}, Eq.~\eqref{eq:fragility}),
and $T_{g}$ is the glass transition temperature defined as the temperature
were viscosity is \SI{e12}{Pa.s}.

\begin{multline}
\log_{10}\left(\eta\left(T,\eta_{\infty},T_{g},m\right)\right)=\log_{10}\left(\eta_{\infty}\right)\\
+\frac{T_{g}}{T}\left(12-\log_{10}\left(\eta_{\infty}\right)\right)\\
\times\exp\left(\left(\frac{T_{g}}{T}-1\right)\left(\frac{m}{12-\log_{10}\left(\eta_{\infty}\right)}-1\right)\right)\label{eq:MYEGA}
\end{multline}

\begin{equation}
m\equiv\left.\frac{\partial\log_{10}\left(\eta\left(T\right)\right)}{\partial\left(T_{g}/T\right)}\right|_{T=T_{g}}\label{eq:fragility}
\end{equation}

This work aimed to develop and test a reproducible gray-box NN to
predict the tem\-per\-a\-ture-dependence of viscosity. This work
includes (a) a pre-processing operation with a chemical feature extractor
and a normalization unit, (b) an extended chemical domain of 39 chemical
compounds, and (c) a permissive license that allows the community
to use and improve both data and code (see Section \ref{subsec:Availability-and-reproducibility}).

\section{Materials and methods}

\subsection{Data collection and preparation \label{subsec:Data-collection}}

Data used in this work come from the SciGlass database, which is publicly
available under the Open Database License (\url{https://github.com/epam/SciGlass}).
This work focused on oxide liquids, which are the majority of the
available data in SciGlass. Data points with viscosity greater than
\SI{e12}{Pa.s} were discarded, as these measurements have a higher
probability of being underestimated due to the long times required
to reach equilibrium. Data points with viscosity smaller than \SI{e-5}{Pa.s}
were also discarded, as such low viscosity is probably due to measurement
error. A deduplication routine was then applied to the dataset by
following three steps:
\begin{enumerate}
\item rounding the chemical composition (in mole fraction) to the 2\textsuperscript{nd}
decimal place, and the temperature (in Kelvin) to the closest integer;
\item grouping the examples with the same chemical composition and temperature;
\item taking the median value of the base-10 logarithm of viscosity for
each group, thus creating a new dataset with only one example per
group.
\end{enumerate}
The next step was the cleaning process. Each liquid in the deduplicated
dataset was analyzed individually and had to meet the following criteria:
at least 3 data points with $\eta\geq$~\SI{e7}{Pa.s} and at least
3 data points with $\eta\leq$~\SI{e4}{Pa.s}. The rationale is to
guarantee a minimum amount of ``high'' and ``low'' viscosity data
points in the hope that this ``holistic'' view of the phenomenon
improves the prediction power of the induced model. Liquids that did
not meet these criteria were not considered in this work.

The cleaning stage also addressed the presence of outliers, which
can impact the predictive power of the model. To identify these outliers,
a non-linear regression of the MYEGA equation (Eq.~\eqref{eq:MYEGA})
was performed using the temperature and viscosity data points for
each liquid individually. This process was performed using least-squares
with a smooth $L_{1}$ loss function (robust to outliers) and the
Trust Region Reflective algorithm \citep{branch1999subspace}. Data
points with a residual greater than or equal to one were labeled as
outliers and discarded; a new regression of the MYEGA equation was
performed in these cases.

Some well-studied liquids such as \ce{SiO2} and \ce{B2O3} have viscosity
datasets with a significant variance, which can impact the predictive
power of the model. Liquids with high variance were discarded by only
considering datasets with a cost of regression of the MYEGA equation
lower than 7. This threshold was selected by visual analysis of all
the viscosity datasets and respective regression.

The final cleaning step considered the viscosity function parameters
obtained by the non-linear regression of the MYEGA equation. Only
liquids with $\eta_{\infty}<$~\SI{e5}{Pa.s} and $10<m<120$ and
$T_{g}>$~\SI{300}{K} where considered in this work. These are valid
inequalities for these parameters considering the available knowledge
on oxide liquids.

The dataset contained \num{17584} data points of 847 different oxide
liquids after cleaning. From this dataset, 85 liquids were randomly
selected, and their data points were collected into the test dataset.
The test dataset was \emph{not} used for hyperparameter tuning (see
Section \ref{subsec:Hyperparameters}), and it was \emph{not} used
for training the neural network; it was only used to access the predictive
power of the final trained model. 

All calculations were performed in the base-10 logarithmic scale of
viscosity due to the immense difference between the lowest and highest
viscosity values in this dataset (14 orders of magnitude).

\subsection{Machine learning pipeline \label{subsec:Pipeline_Overview}}

\begin{figure*}[t]
\begin{centering}
\includegraphics[width=0.8\textwidth]{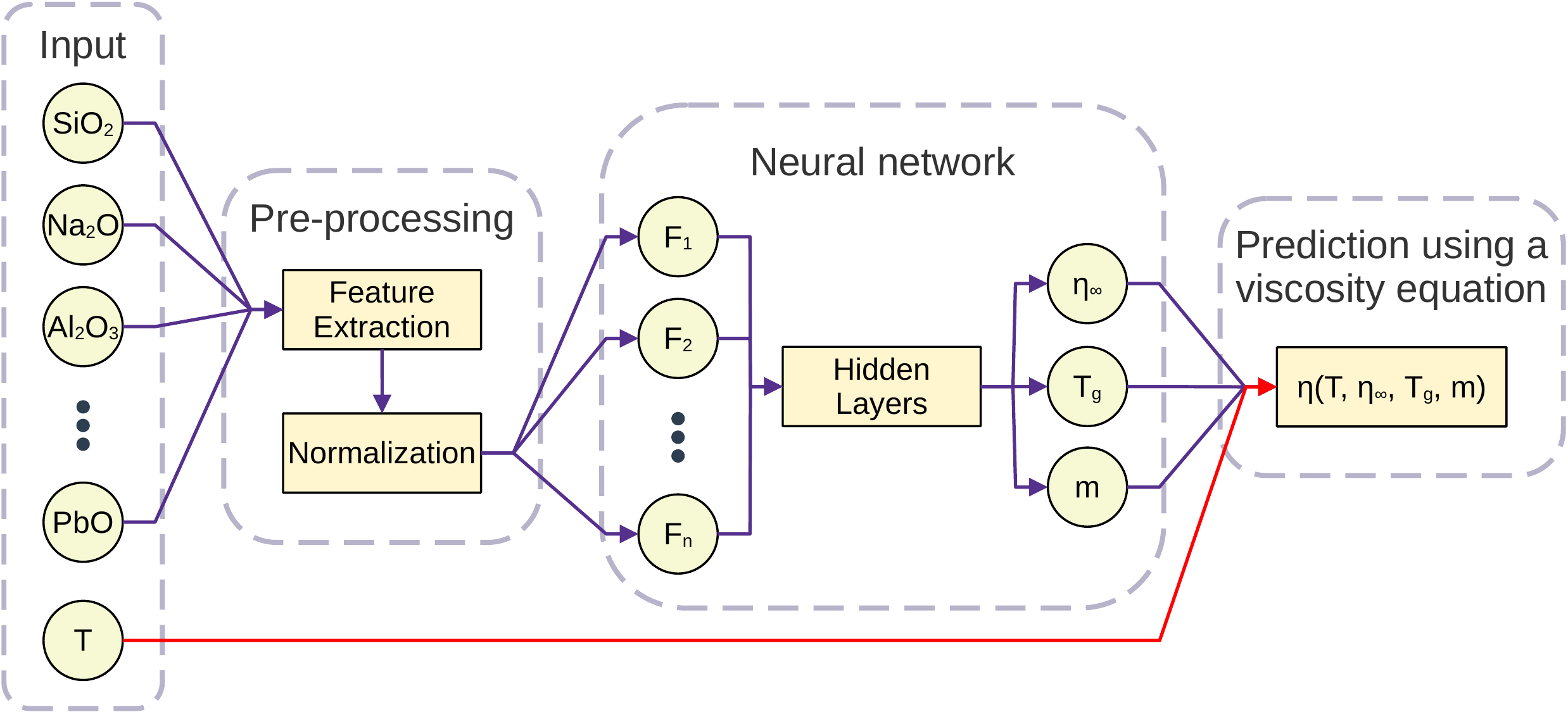}
\par\end{centering}
\caption{Flowchart of the machine learning pipeline used in this work. \label{fig:Schematics-pipeline}}
\end{figure*}

\emph{Neural network} is a general term for a group of machine learning
algorithms used for pattern recognition, which is performed by an
assortment of interconnected computational units called neurons. In
materials science, NNs can be applied in many types of problems and
are often used for their universal regressor capabilities. This work
focuses on feedforward multilayer perceptron NNs, one of the most
simple architectures of NNs. For more information about the mathematical
and statistical basis of this topic, see Ref.~\citep{aggarwal2018neural}.

This work was inspired by the gray-box NN recently published by Tandia
et al.~\citep{tandia2019machine}. A new item proposed and tested
here is an additional step in the pipeline: a pre-processing step
that includes a feature extractor and a scaler. The feature extractor
will be described in Section \ref{subsec:Feature-extractor}; the
scaler is a unit that computes the z-score of the features supplied
to the NN to reduce the bias of those features with a higher magnitude
(see the \apdx{} for more information on the z-score).

Figure \ref{fig:Schematics-pipeline} shows a flowchart of the machine
learning pipeline used here. The arrows indicate the flow of information
that starts from the input data (composition and temperature of the
liquid) and ends with the prediction of viscosity by the viscosity
equation.

The neural network in the pipeline is a predictor of the viscosity
parameters, namely $\eta_{\infty}$, $T_{g}$, and $m$. These parameters,
along with temperature, serve as inputs for the viscosity equation,
which makes the final prediction. One advantage of this gray-box approach\textemdash in
contrast with a black-box approach\textemdash is that the viscosity
parameters can be predicted individually. Hence, the same trained
model predicts not only the tem\-per\-a\-ture-dependence of viscosity
but also the liquid's fragility index and its glass transition temperature.

The machine learning pipeline shown in Fig.~\ref{fig:Schematics-pipeline}
has many hyperparameters, such as the number of hidden layers in the
NN and their size, for example. The methodology for finding a good
set of hyperparameters is discussed in detail in Section \ref{subsec:Hyperparameters}.
However, some hyperparameters were fixed from the beginning as design
choices. The viscosity equation used in the pipeline is one of these
fixed hyperparameters; it was the MYEGA equation. The backpropagation
loss function is another fixed hyperparameter; it was chosen as the
mean squared error (MSE) because it is a suitable loss function to
solve regression problems.

\subsection{Feature extraction and selection\label{subsec:Feature-extractor}}

\begin{table}
\caption{Chemical properties considered in this work. \textsuperscript{\textdagger}
based on DFT volume of the OQMD ground state \citep{saal2013materials,kirklin2015open}.
\protect\textsuperscript{{*}} computed in DFT simulation of $T=$~\SI{0}{K}
ground state. \label{tab:Chemical-features}}

\selectlanguage{brazil}%
\centering{}\resizebox{!}{.47\textheight}{\foreignlanguage{english}{}%
\begin{tabular}{l}
\toprule 
\selectlanguage{english}%
Atomic number\selectlanguage{brazil}%
\tabularnewline
\selectlanguage{english}%
Atomic weight\selectlanguage{brazil}%
\tabularnewline
\selectlanguage{english}%
Atomic volume\selectlanguage{brazil}%
\tabularnewline
\selectlanguage{english}%
Atomic radius \citep{slater1964atomic}\selectlanguage{brazil}%
\tabularnewline
\selectlanguage{english}%
Atomic radius \citep{rahm2016atomic,rahm2017corrigendum}\selectlanguage{brazil}%
\tabularnewline
\selectlanguage{english}%
Boiling point\selectlanguage{brazil}%
\tabularnewline
\selectlanguage{english}%
Melting point\selectlanguage{brazil}%
\tabularnewline
\selectlanguage{english}%
C\textsubscript{6} coefficient \citep{gould2016c6}\selectlanguage{brazil}%
\tabularnewline
\selectlanguage{english}%
Covalent radius \citep{cordero2008covalent}\selectlanguage{brazil}%
\tabularnewline
\selectlanguage{english}%
Single-bond covalent radius \citep{pyykko2009molecular}\selectlanguage{brazil}%
\tabularnewline
\selectlanguage{english}%
Double-bond covalent radius \citep{pyykko2009molecular-1}\selectlanguage{brazil}%
\tabularnewline
\selectlanguage{english}%
Density\selectlanguage{brazil}%
\tabularnewline
\selectlanguage{english}%
Dipole polarizability\selectlanguage{brazil}%
\tabularnewline
\selectlanguage{english}%
Effective nuclear charge\selectlanguage{brazil}%
\tabularnewline
\selectlanguage{english}%
Electron affinity\selectlanguage{brazil}%
\tabularnewline
\selectlanguage{english}%
Energy to remove the first electron\selectlanguage{brazil}%
\tabularnewline
\selectlanguage{english}%
Electronegativity in the Gosh scale \citep{ghosh2005new}\selectlanguage{brazil}%
\tabularnewline
\selectlanguage{english}%
Electronegativity in the Pauling scale\selectlanguage{brazil}%
\tabularnewline
\selectlanguage{english}%
Electronegativity in the Sanderson scale\selectlanguage{brazil}%
\tabularnewline
\selectlanguage{english}%
Electronegativity in the Martynov-Batsanov scale\selectlanguage{brazil}%
\tabularnewline
\selectlanguage{english}%
Fusion enthalpy\selectlanguage{brazil}%
\tabularnewline
\selectlanguage{english}%
Glawe's number \citep{glawe2016optimal}\selectlanguage{brazil}%
\tabularnewline
\selectlanguage{english}%
Mendeleev's number\selectlanguage{brazil}%
\tabularnewline
\selectlanguage{english}%
Pettifor's number \citep{pettifor1984chemical}\selectlanguage{brazil}%
\tabularnewline
\selectlanguage{english}%
Heat of formation\selectlanguage{brazil}%
\tabularnewline
\selectlanguage{english}%
Lattice constant\selectlanguage{brazil}%
\tabularnewline
\selectlanguage{english}%
BCC lattice parameter\textsuperscript{\textdagger}\selectlanguage{brazil}%
\tabularnewline
\selectlanguage{english}%
FCC lattice parameter\textsuperscript{\textdagger}\selectlanguage{brazil}%
\tabularnewline
\selectlanguage{english}%
Mass number of the most abundant isotope\selectlanguage{brazil}%
\tabularnewline
\selectlanguage{english}%
Maximum ionization energy\selectlanguage{brazil}%
\tabularnewline
\selectlanguage{english}%
Number of electrons\selectlanguage{brazil}%
\tabularnewline
\selectlanguage{english}%
Number of neutrons\selectlanguage{brazil}%
\tabularnewline
\selectlanguage{english}%
Number of protons\selectlanguage{brazil}%
\tabularnewline
\selectlanguage{english}%
Number of valence electrons \citep{mentel2020mendeleev}\selectlanguage{brazil}%
\tabularnewline
\selectlanguage{english}%
Number of valence electrons \citep{ward2018matminer}\selectlanguage{brazil}%
\tabularnewline
\selectlanguage{english}%
Number of unfilled valence orbitals\selectlanguage{brazil}%
\tabularnewline
\selectlanguage{english}%
Number of unfilled s valence orbitals\selectlanguage{brazil}%
\tabularnewline
\selectlanguage{english}%
Number of unfilled p valence orbitals\selectlanguage{brazil}%
\tabularnewline
\selectlanguage{english}%
Number of unfilled d valence orbitals\selectlanguage{brazil}%
\tabularnewline
\selectlanguage{english}%
Number of unfilled f valence orbitals\selectlanguage{brazil}%
\tabularnewline
\selectlanguage{english}%
Number of filled s valence orbitals\selectlanguage{brazil}%
\tabularnewline
\selectlanguage{english}%
Number of filled p valence orbitals\selectlanguage{brazil}%
\tabularnewline
\selectlanguage{english}%
Number of filled d valence orbitals\selectlanguage{brazil}%
\tabularnewline
\selectlanguage{english}%
Number of filled f valence orbitals\selectlanguage{brazil}%
\tabularnewline
\selectlanguage{english}%
Number of oxidation states\selectlanguage{brazil}%
\tabularnewline
\selectlanguage{english}%
Bandgap energy\textsuperscript{{*}}\selectlanguage{brazil}%
\tabularnewline
\selectlanguage{english}%
Energy per atom\textsuperscript{{*}}\selectlanguage{brazil}%
\tabularnewline
\selectlanguage{english}%
Magnetic moment\textsuperscript{{*}}\selectlanguage{brazil}%
\tabularnewline
\selectlanguage{english}%
Volume per atom\textsuperscript{{*}}\selectlanguage{brazil}%
\tabularnewline
\selectlanguage{english}%
Space group\textsuperscript{{*}}\selectlanguage{brazil}%
\tabularnewline
\selectlanguage{english}%
Radii of element in metallic glass structure\selectlanguage{brazil}%
\tabularnewline
\selectlanguage{english}%
Van der Walls radius \citep{haynes2014crc}\selectlanguage{brazil}%
\tabularnewline
\selectlanguage{english}%
Van der Walls radius \citep{alvarez2013cartography}\selectlanguage{brazil}%
\tabularnewline
\selectlanguage{english}%
Van der Walls radius \citep{batsanov2001van}\selectlanguage{brazil}%
\tabularnewline
\selectlanguage{english}%
Van der Walls radius \citep{rappe1992uff}\selectlanguage{brazil}%
\tabularnewline
\selectlanguage{english}%
Van der Walls radius \citep{allinger1994molecular}\selectlanguage{brazil}%
\tabularnewline
\bottomrule
\end{tabular}\foreignlanguage{english}{}}\selectlanguage{english}%
\end{table}

In the \emph{chemical composition} domain, the features of a liquid
are represented by a vector with the atomic mole fraction of its constituents.
To convert these features to the \emph{chemical property} domain,
one must choose a chemical property and an aggregator function. An
example is to choose the atomic weight as a property and compute the
``mean atomic weight,'' which is a feature of the liquid. By choosing
different chemical properties and aggregator functions, one can ``extract''
new features from the liquid in the chemical property domain. This
procedure is called feature extraction or feature engineering \citep{ward2016generalpurpose}. 

The mathematical procedure for this process starts by creating the
vector $\boldsymbol{C}=\left[x_{1},x_{2},\ldots,x_{n}\right]$ of
the atomic mole fractions of the chemical elements $e_{1}$, $e_{2}$,
\ldots , $e_{n}$ that make a certain liquid. Let $\boldsymbol{S}=\left[s_{1},s_{2},\ldots,s_{n}\right]$
be the vector of a certain chemical property $s_{i}$ of the chemical
element $e_{i}$ (the atomic weight, for example). We compute the
property vectors $\boldsymbol{W}$ (weighted) and $\boldsymbol{A}$
(absolute) as 

\begin{equation}
\boldsymbol{W}=\boldsymbol{C}\boldsymbol{S}^{T},\label{eq:feat_vec_weighted}
\end{equation}
and

\begin{equation}
\boldsymbol{A}=\left\lceil \boldsymbol{C}\right\rceil \boldsymbol{S}^{T}.\label{eq:feat_vec_absolute}
\end{equation}
Note that the ceil function is applied element-wise in vector $\boldsymbol{C}$
in Eq.~\eqref{eq:feat_vec_absolute}.

Finally, by applying an aggregator function to the items of the vectors
$\boldsymbol{W}$ or $\boldsymbol{A}$, one obtains a particular chemical
feature of the liquid. The aggregator functions used in this work
are summation, mean, standard deviation, minimum, and maximum. Many
features can be extracted following this procedure. This work considered
all the chemical properties shown in Table \ref{tab:Chemical-features},
which are available in the Python modules \texttt{mendeleev} \citep{mentel2020mendeleev}
and \texttt{matminer} \citep{ward2018matminer}. 

A total of 601 chemical property features were extracted using this
procedure. A feature selection routine to eliminate features with
high collinearity and low variance was performed as follows:
\begin{enumerate}
\item Let $l$ be the set of all chemical property features;
\item Compute the variance inflation factor (VIF) \citep{marquardt1970generalized}
of all the features in $l$;\label{enu:feature_selection_1}
\item Let $v$ be the maximum value of all the computed VIFs;
\item If $v>10$, then the feature associated with this VIF is removed from
$l$ because its collinearity is too high. If a feature was removed
in this step, return to step \ref{enu:feature_selection_1};
\item If $v\leq10$, then the standard deviation is computed for all remaining
features. Those features with a standard deviation of less than \num{e-3}
are removed from $l$ because they have too low variance;
\item The remaining features in $l$ are the features used in this work.
\end{enumerate}
Only chemical property features were extracted in this work. However,
in the framework proposed by Adam and Gibbs \citep{adam1965temperature}
(which is the basis for the MYEGA equation), viscosity depends on
the size of the cooperative rearranging regions, which are related
to the atomic structure of the liquid. Therefore, a predictor of viscosity
that uses only chemical features is unlikely to generalize \emph{all}
the intricacies of viscous flow. Structural features, however, are
outside of the scope of this work. To avoid data leakage \citep{kaufman2012leakage},
feature selection was performed using only data \emph{not} reserved
in the test dataset.

\subsection{Hyperparameter tuning \label{subsec:Hyperparameters}}

The prediction power and the generalization power of a neural network
are highly dependent on its hyperparameters (HP), such as the number
of neurons, number of hidden layers, and activation function. Determining
a good set of HP for a new problem is not trivial. Thus, before settling
for the final network architecture, it is vital to test many sets
of HP, a process called \emph{hyperparameter tuning}.

HP tuning was performed in a sequence of three experiments, starting
with a random search, then a Bayesian search, and finally, a 10-fold
cross-validation. These experiments were done in series, with the
second and third using knowledge obtained in the previous experiments.

The \emph{first} experiment was the test of 500 different HP sets,
randomly drawn from the search space shown in Table \ref{tab:Seach-spaces-and-best-architecture}.
For each HP set that was drawn, a neural network was built and trained.
Only the 762 liquids that were not part of the test dataset were used
in this experiment. The training and validation datasets were the
same for all NNs in this experiment and consisted of the data points
of 686 and 76 randomly chosen liquids, respectively (90\textendash 10
split). This split strategy was chosen with extrapolation (instead
of interpolation) performance in mind: all the data points in the
validation dataset are from liquids with different chemical composition
than those in the training dataset. The training of the NNs was terminated
if their performance was not good enough when compared with the finished
trials (using an Asynchronous Successive Halving Algorithm, ASHA \citep{li2020ai}),
or until no improvement in the prediction of the validation dataset
was observed after a particular number of epochs determined by the
``patience'' hyperparameter. The average MSE loss of the validation
dataset was recorded for all HP sets.

\begin{table*}[!t]
\caption{Hyperparameters search space for experiments 1 and 2, and selected
HP set after cross-validation. The functions Tanh and ReLU are the
hyperbolic tangent and the rectifier linear unit. SGD is the stochastic
gradient descent \citep{kiefer1952stochastic,robbins1951stochastic}.
Adam \citep{kingma2017adam} and AdamW \citep{kingma2017adam,loshchilov2019decoupled}
are stochastic gradient descent methods based on an adaptive estimation
of first-order and second-order moments, the latter having a weight
decay coefficient. \label{tab:Seach-spaces-and-best-architecture}}

\centering{}%
\begin{tabular}{lccc}
\toprule 
Hyperparameter & 1\textsuperscript{st} Experiment & 2\textsuperscript{nd} Experiment & Selected\tabularnewline
\midrule
\midrule 
Number of hidden layers & \multicolumn{2}{c}{\{1,2,3\}} & 2\tabularnewline
Training batch size & \multicolumn{2}{c}{\{2, 4, 8, 16, 32, 64, 128\}} & 64\tabularnewline
Patience (integer) & {[}5, 20{]} & {[}5, 25{]} & 9\tabularnewline
Optimizer & \multicolumn{2}{c}{\{SGD, Adam, AdamW\}} & AdamW\tabularnewline
Optimizer learning rate & \multicolumn{2}{c}{{[}\num{e-5}, \num{e-1}{]}} & \num{1.16e-3}\tabularnewline
SGD momentum & \multicolumn{2}{c}{{[}0, 1{]}} & \textemdash{}\tabularnewline
\vspace{-10bp}
 &  &  & \tabularnewline
1\textsuperscript{st} hidden layer &  &  & \tabularnewline
\midrule
\enspace{}Size (integer) & {[}16, 256{]} & {[}16, 512{]} & 192\tabularnewline
\enspace{}Dropout (\%) & \multicolumn{2}{c}{{[}0, 50{]}} & 7.94\tabularnewline
\enspace{}Use batch normalization & \multicolumn{2}{c}{\{Yes, No\}} & No\tabularnewline
\enspace{}Activation function & \multicolumn{2}{c}{\{ReLU, Tanh\}} & ReLU\tabularnewline
\vspace{-10bp}
 &  &  & \tabularnewline
2\textsuperscript{nd} hidden layer &  &  & \tabularnewline
\midrule
\enspace{}Size (integer) & \multicolumn{2}{c}{{[}16, 256{]}} & 48\tabularnewline
\enspace{}Dropout (\%) & \multicolumn{2}{c}{{[}0, 50{]}} & 5.37\tabularnewline
\enspace{}Use batch normalization & \multicolumn{2}{c}{\{Yes, No\}} & No\tabularnewline
\enspace{}Activation function & \multicolumn{2}{c}{\{ReLU, Tanh\}} & Tanh\tabularnewline
\vspace{-10bp}
 &  &  & \tabularnewline
3\textsuperscript{rd} hidden layer &  &  & \tabularnewline
\midrule
\enspace{}Size (integer) & \multicolumn{2}{c}{{[}16, 256{]}} & \textemdash{}\tabularnewline
\enspace{}Dropout (\%) & \multicolumn{2}{c}{{[}0, 50{]}} & \textemdash{}\tabularnewline
\enspace{}Use batch normalization & \multicolumn{2}{c}{\{Yes, No\}} & \textemdash{}\tabularnewline
\enspace{}Activation function & \multicolumn{2}{c}{\{ReLU, Tanh\}} & \textemdash{}\tabularnewline
\bottomrule
\end{tabular}
\end{table*}

The \emph{second} experiment was the test of 300 different HP sets,
drawn from the search space shown in Table \ref{tab:Seach-spaces-and-best-architecture}.
This search space is almost identical to that of the first experiment,
but allowing for bigger size of the first hidden layer and for higher
values of patience\textemdash a choice made after observing that the
top 100 HP sets from the previous experiment were too close to the
upper bound limit of these two hyperparameters. All the other characteristics
of the second experiment are the same as the first, except that the
HP sets were not drawn randomly but instead guided by a Tree-struc\-tured
Parzen Estimator algorithm \citep{bergstra2011algorithms}. The first
20 HP sets, however, were the 20 HP sets that performed the best in
the first experiment, that is, those with the lowest average MSE loss
of the validation dataset.

The \emph{third} and final experiment was a 10-fold cross-validation
for each of the 10 HP sets with the lowest average MSE validation
loss among all the 700 HP sets tested in this work. A 10-fold cross-validation
experiment consists of splitting the data into 10 different sets called
folds. Each of these folds is selected once to be the validation dataset,
with the remaining folds making the training\linebreak{}
dataset. A NN is trained for each of these different training and
validation datasets. The HP set with the lowest average of the validation
losses in this analysis was selected as the final architecture for
the neural network. For all three experiments, the validation dataset
always contained only liquids that were not present in the training
dataset.

All experiments were coded in the Python programming language, and
the NN were trained using a personal computer with an 8-core CPU and
16 GB of RAM. The neural networks were built using the \texttt{PyTorch-Lightning}
module \citep{falcon2020pytorchlightning}, a high-level interface
for \texttt{PyTorch} \citep{paszke2019pytorch}. Data management was
performed with the \texttt{pandas} module \citep{mckinney2010data}.
Hyperparameter tuning was performed using \texttt{tune} \citep{liaw2018tune}
(first and second experiments) and \texttt{hyperopt} \citep{bergstra2013making}
(second experiment). Cross-validation and data splitting was performed
using \texttt{sklearn} \citep{pedregosa2011scikitlearn}.

\section{Results and discussion}

\subsection{Data analysis}

\begin{figure*}[!t]
\begin{centering}
\subfloat[\label{fig:dataset_num_compounds}]{\begin{centering}
\includegraphics[width=0.45\textwidth]{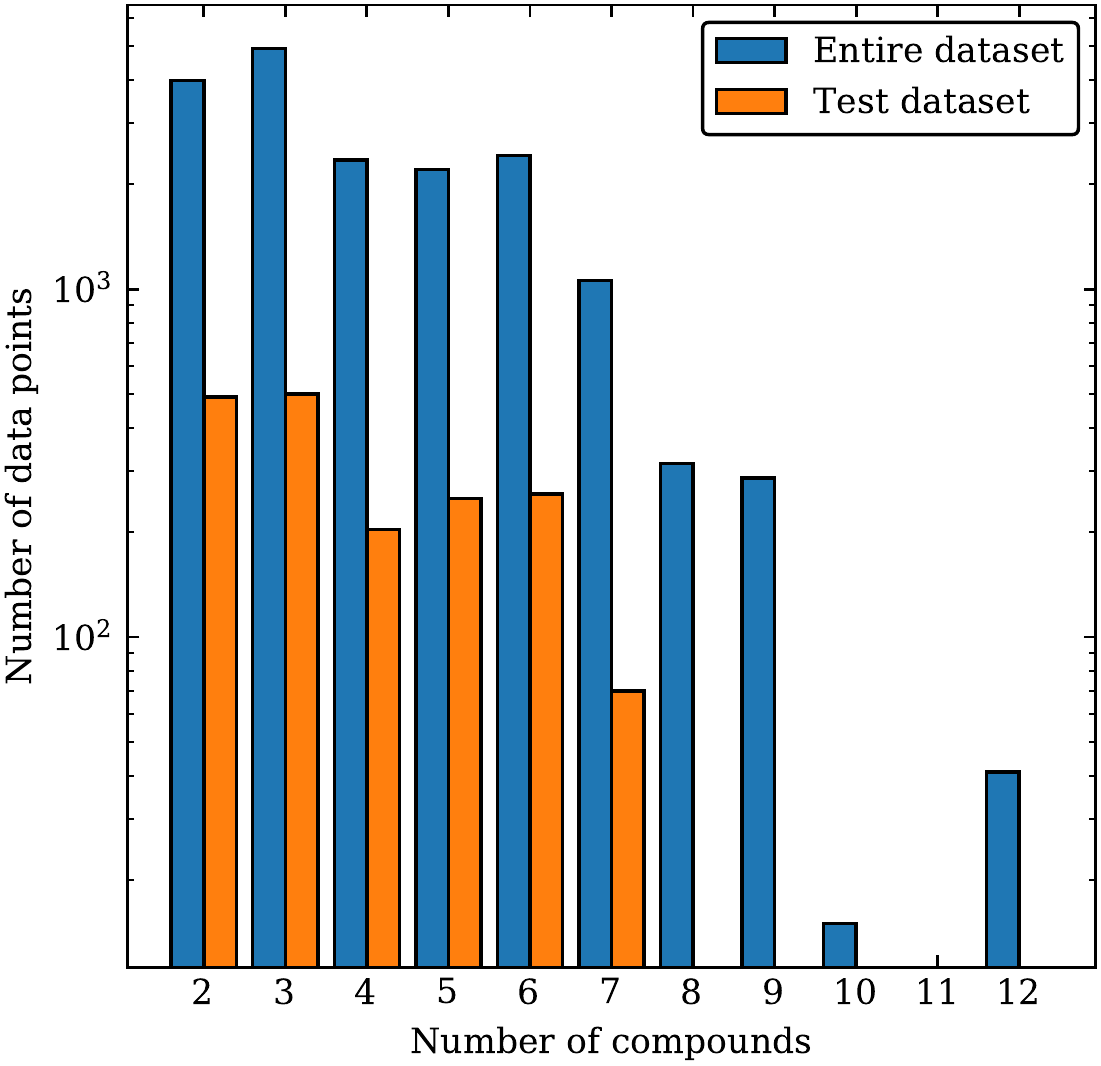}
\par\end{centering}
}\subfloat[\label{fig:dataset_visc_vals}]{\begin{centering}
\includegraphics[width=0.45\textwidth]{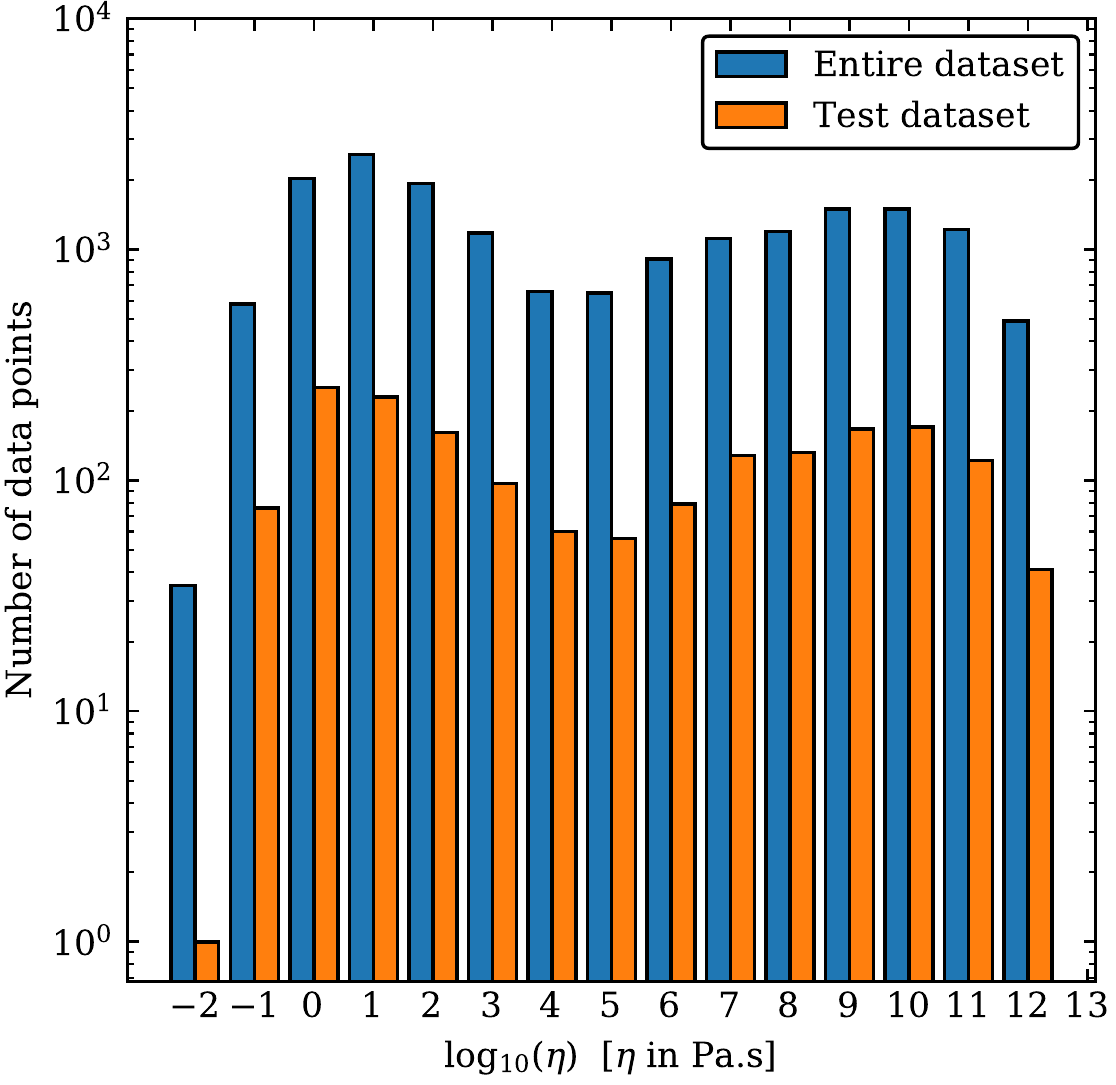}
\par\end{centering}
}
\par\end{centering}
\begin{centering}
\subfloat[\label{fig:dataset_compounds}]{\begin{centering}
\includegraphics[width=0.9\textwidth]{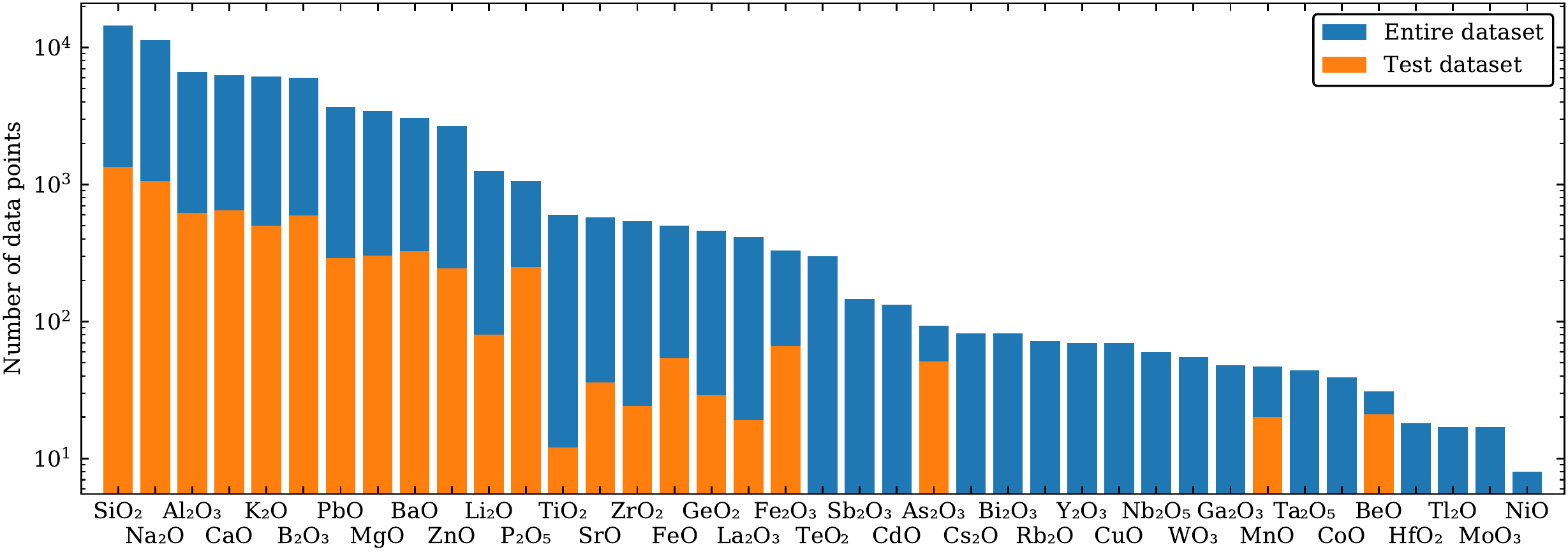}
\par\end{centering}
}
\par\end{centering}
\caption{Data analysis of the entire dataset and the test dataset. (\textbf{a})
Histogram of the number of compounds in the liquid. (\textbf{b}) Histogram
of the viscosity values. (\textbf{c}) Histogram for each compound,
the orange regions represent the fraction selected for the test dataset.
\label{fig:dataset-analysis}}
\end{figure*}

Figure \ref{fig:dataset-analysis} has three plots with information
on the entire viscosity dataset used in this work, together with the
test dataset. The plot in Fig.~\ref{fig:dataset_num_compounds} shows
the histogram of the number of different compounds of the liquids,
the overall maximum is 12, and the maximum of the test dataset is
7. As expected, the number of data points has an inverse correlation
with the number of compounds because simple liquids are more studied
than complex ones.

The histogram in Fig.~\ref{fig:dataset_visc_vals} shows a bimodal
distribution of the viscosity values, with a local minimum around
the center at \SI{e5}{Pa.s}. Measuring viscosity in this central
region is challenging: the liquid has enough kinetic enengy and thermodynamic
driving force to crystallize, which often happens too fast and forcibly
ends the viscometry experiment.

Finally, the histogram in Fig.~\ref{fig:dataset_compounds} shows
the number of data points per compound for the entire dataset, with
the fraction used for the test dataset marked in orange. Most liquids
are made with \ce{SiO2}, \ce{Na2O}, and \ce{Al2O3}, which are common
compounds used in the glass industry. Not all compounds are part of
the test dataset because of the way it was produced. As discussed
in Section \ref{subsec:Data-collection}, the test dataset contains
85 randomly selected liquids from the entire dataset, not a certain
number of randomly selected data points from this dataset. 

\subsection{Feature extraction and selection}

The feature extraction procedure generated a total of 601 chemical
property features extracted from the chemical composition. From this
total, 35 features shown in Table \ref{tab:selected-features} were
selected. The feature selection procedure considered the collinearity
and variance of the 601 initial features, not their relationship with
viscosity. Finding which of these features are more or less relevant
to model viscosity was a task left to the NN.

\begin{table}[!t]
\caption{Type, aggregator function, and chemical property of the 35 selected
features. \textquotedblleft W\textquotedblright{} stands for weighed
(see Eq.~\eqref{eq:feat_vec_weighted}) and \textquotedblleft A\textquotedblright{}
stands for absolute (see Eq.~\eqref{eq:feat_vec_absolute}). SD is
the standard deviation. \label{tab:selected-features}}

\selectlanguage{brazil}%
\centering{}\resizebox{\columnwidth}{!}{
\begin{tabular}{lll}
\toprule 
\selectlanguage{english}%
Type\selectlanguage{brazil}%
 & \selectlanguage{english}%
Aggregator\selectlanguage{brazil}%
 & \selectlanguage{english}%
Chemical property\selectlanguage{brazil}%
\tabularnewline
\midrule
\midrule 
\selectlanguage{english}%
W\selectlanguage{brazil}%
 & \selectlanguage{english}%
Maximum\selectlanguage{brazil}%
 & \selectlanguage{english}%
Atomic radius\selectlanguage{brazil}%
\tabularnewline
\selectlanguage{english}%
W\selectlanguage{brazil}%
 & \selectlanguage{english}%
Maximum\selectlanguage{brazil}%
 & \selectlanguage{english}%
Atomic volume\selectlanguage{brazil}%
\tabularnewline
\selectlanguage{english}%
W\selectlanguage{brazil}%
 & \selectlanguage{english}%
Maximum\selectlanguage{brazil}%
 & \selectlanguage{english}%
Bandgap energy\selectlanguage{brazil}%
\tabularnewline
\selectlanguage{english}%
W\selectlanguage{brazil}%
 & \selectlanguage{english}%
Maximum\selectlanguage{brazil}%
 & \selectlanguage{english}%
C\textsubscript{6} coefficient \citep{gould2016c6}\selectlanguage{brazil}%
\tabularnewline
\selectlanguage{english}%
W\selectlanguage{brazil}%
 & \selectlanguage{english}%
Maximum\selectlanguage{brazil}%
 & \selectlanguage{english}%
Number of filled d valence orbitals\selectlanguage{brazil}%
\tabularnewline
\selectlanguage{english}%
W\selectlanguage{brazil}%
 & \selectlanguage{english}%
Maximum\selectlanguage{brazil}%
 & \selectlanguage{english}%
Number of unfilled s orbitals\selectlanguage{brazil}%
\tabularnewline
\selectlanguage{english}%
W\selectlanguage{brazil}%
 & \selectlanguage{english}%
Maximum\selectlanguage{brazil}%
 & \selectlanguage{english}%
Number of unfilled d orbitals\selectlanguage{brazil}%
\tabularnewline
\selectlanguage{english}%
W\selectlanguage{brazil}%
 & \selectlanguage{english}%
Maximum\selectlanguage{brazil}%
 & \selectlanguage{english}%
Number of oxidation states\selectlanguage{brazil}%
\tabularnewline
\selectlanguage{english}%
W\selectlanguage{brazil}%
 & \selectlanguage{english}%
Maximum\selectlanguage{brazil}%
 & \selectlanguage{english}%
Number of valence electrons \citep{ward2018matminer}\selectlanguage{brazil}%
\tabularnewline
\selectlanguage{english}%
W\selectlanguage{brazil}%
 & \selectlanguage{english}%
Maximum\selectlanguage{brazil}%
 & \selectlanguage{english}%
Space group\selectlanguage{brazil}%
\tabularnewline
\selectlanguage{english}%
W\selectlanguage{brazil}%
 & \selectlanguage{english}%
Maximum\selectlanguage{brazil}%
 & \selectlanguage{english}%
Volume per atom\selectlanguage{brazil}%
\tabularnewline
\selectlanguage{english}%
W\selectlanguage{brazil}%
 & \selectlanguage{english}%
Minimum\selectlanguage{brazil}%
 & \selectlanguage{english}%
Fusion enthalpy\selectlanguage{brazil}%
\tabularnewline
\selectlanguage{english}%
W\selectlanguage{brazil}%
 & \selectlanguage{english}%
Minimum\selectlanguage{brazil}%
 & \selectlanguage{english}%
C\textsubscript{6} coefficient \citep{gould2016c6}\selectlanguage{brazil}%
\tabularnewline
\selectlanguage{english}%
W\selectlanguage{brazil}%
 & \selectlanguage{english}%
Minimum\selectlanguage{brazil}%
 & \selectlanguage{english}%
Maximum ionization energy\selectlanguage{brazil}%
\tabularnewline
\selectlanguage{english}%
W\selectlanguage{brazil}%
 & \selectlanguage{english}%
Minimum\selectlanguage{brazil}%
 & \selectlanguage{english}%
Number of valence electrons \citep{ward2018matminer}\selectlanguage{brazil}%
\tabularnewline
\selectlanguage{english}%
W\selectlanguage{brazil}%
 & \selectlanguage{english}%
Minimum\selectlanguage{brazil}%
 & \selectlanguage{english}%
Number of valence electrons \citep{mentel2020mendeleev}\selectlanguage{brazil}%
\tabularnewline
\selectlanguage{english}%
W\selectlanguage{brazil}%
 & \selectlanguage{english}%
Minimum\selectlanguage{brazil}%
 & \selectlanguage{english}%
Space group\selectlanguage{brazil}%
\tabularnewline
\selectlanguage{english}%
W\selectlanguage{brazil}%
 & \selectlanguage{english}%
Mean\selectlanguage{brazil}%
 & \selectlanguage{english}%
Magnetic moment\selectlanguage{brazil}%
\tabularnewline
\selectlanguage{english}%
W\selectlanguage{brazil}%
 & \selectlanguage{english}%
SD\selectlanguage{brazil}%
 & \selectlanguage{english}%
Radii of element in metallic glass structure\selectlanguage{brazil}%
\tabularnewline
\selectlanguage{english}%
A\selectlanguage{brazil}%
 & \selectlanguage{english}%
SD\selectlanguage{brazil}%
 & \selectlanguage{english}%
Atomic radius \citep{rahm2016atomic,rahm2017corrigendum}\selectlanguage{brazil}%
\tabularnewline
\selectlanguage{english}%
A\selectlanguage{brazil}%
 & \selectlanguage{english}%
SD\selectlanguage{brazil}%
 & \selectlanguage{english}%
C\textsubscript{6} coefficient \citep{gould2016c6}\selectlanguage{brazil}%
\tabularnewline
\selectlanguage{english}%
A\selectlanguage{brazil}%
 & \selectlanguage{english}%
SD\selectlanguage{brazil}%
 & \selectlanguage{english}%
Effective nuclear charge\selectlanguage{brazil}%
\tabularnewline
\selectlanguage{english}%
A\selectlanguage{brazil}%
 & \selectlanguage{english}%
SD\selectlanguage{brazil}%
 & \selectlanguage{english}%
Electron affinity\selectlanguage{brazil}%
\tabularnewline
\selectlanguage{english}%
A\selectlanguage{brazil}%
 & \selectlanguage{english}%
SD\selectlanguage{brazil}%
 & \selectlanguage{english}%
Energy per atom\selectlanguage{brazil}%
\tabularnewline
\selectlanguage{english}%
A\selectlanguage{brazil}%
 & \selectlanguage{english}%
SD\selectlanguage{brazil}%
 & \selectlanguage{english}%
Fusion enthalpy\selectlanguage{brazil}%
\tabularnewline
\selectlanguage{english}%
A\selectlanguage{brazil}%
 & \selectlanguage{english}%
SD\selectlanguage{brazil}%
 & \selectlanguage{english}%
Lattice constant\selectlanguage{brazil}%
\tabularnewline
\selectlanguage{english}%
A\selectlanguage{brazil}%
 & \selectlanguage{english}%
SD\selectlanguage{brazil}%
 & \selectlanguage{english}%
Magnetic moment\selectlanguage{brazil}%
\tabularnewline
\selectlanguage{english}%
A\selectlanguage{brazil}%
 & \selectlanguage{english}%
SD\selectlanguage{brazil}%
 & \selectlanguage{english}%
Mendeleev's number\selectlanguage{brazil}%
\tabularnewline
\selectlanguage{english}%
A\selectlanguage{brazil}%
 & \selectlanguage{english}%
SD\selectlanguage{brazil}%
 & \selectlanguage{english}%
Number of filled f valence orbitals\selectlanguage{brazil}%
\tabularnewline
\selectlanguage{english}%
A\selectlanguage{brazil}%
 & \selectlanguage{english}%
SD\selectlanguage{brazil}%
 & \selectlanguage{english}%
Number of unfilled p valence orbitals\selectlanguage{brazil}%
\tabularnewline
\selectlanguage{english}%
A\selectlanguage{brazil}%
 & \selectlanguage{english}%
SD\selectlanguage{brazil}%
 & \selectlanguage{english}%
Number of unfilled d valence orbitals\selectlanguage{brazil}%
\tabularnewline
\selectlanguage{english}%
A\selectlanguage{brazil}%
 & \selectlanguage{english}%
SD\selectlanguage{brazil}%
 & \selectlanguage{english}%
Number of oxidation states\selectlanguage{brazil}%
\tabularnewline
\selectlanguage{english}%
A\selectlanguage{brazil}%
 & \selectlanguage{english}%
SD\selectlanguage{brazil}%
 & \selectlanguage{english}%
Number of valence electrons \citep{mentel2020mendeleev}\selectlanguage{brazil}%
\tabularnewline
\selectlanguage{english}%
A\selectlanguage{brazil}%
 & \selectlanguage{english}%
SD\selectlanguage{brazil}%
 & \selectlanguage{english}%
Van der Walls radius \citep{alvarez2013cartography}\selectlanguage{brazil}%
\tabularnewline
\selectlanguage{english}%
A\selectlanguage{brazil}%
 & \selectlanguage{english}%
SD\selectlanguage{brazil}%
 & \selectlanguage{english}%
Van der Walls radius \citep{rappe1992uff}\selectlanguage{brazil}%
\tabularnewline
\bottomrule
\end{tabular}\foreignlanguage{english}{}}\selectlanguage{english}%
\end{table}

\subsection{Building and training the model}

The selected neural network architecture after HP tuning is shown
in the last column of Table \ref{tab:Seach-spaces-and-best-architecture}.
It is a deep network with two hidden layers, the first layer with
192 neurons and the second layer with 48 neurons, both having dropout
\citep{srivastava2014dropout} and not having batch normalization
\citep{ioffe2015batch}. Interestingly, this network has mixed activation
functions, with ReLU for the first layer and Tanh for the second.
The machine learning pipeline containing this network will be called
ViscNet from now on.

Two notable differences between ViscNet and the NN reported by Tandia
et al.~\citep{tandia2019machine} is the number of layers and their
size (number of neurons). Both hyperparameters are higher in this
work; although the final architecture was not disclosed by Tandia
et al., a single-layer with about 10 neurons was strongly suggested.
Nonetheless, the two models differ in scope, considering that the
NN reported by Tandia et al.\ is focused on particular liquid compositions
(9 different chemical compounds). A smaller scope explains the difference
in complexity.

\begin{figure}
\begin{centering}
\includegraphics[width=0.8\columnwidth]{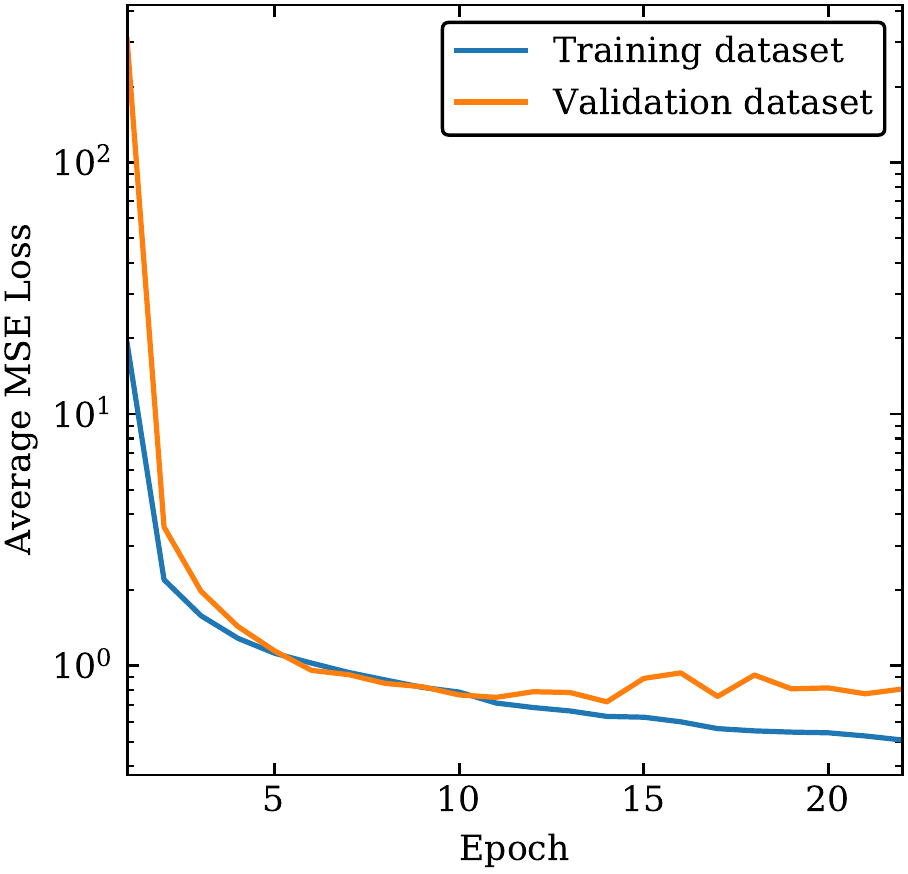}
\par\end{centering}
\caption{Learning curve of the ViscNet neural network. An epoch is when all
the training data passes through the network during the training phase.
\label{fig:ViscNet_learning_curve}}
\end{figure}

Interestingly, the activation function used in by Tandia et al.~\citep{tandia2019machine}
(single-layer) and of the second (final) layer of this work is the
same, a hyperbolic tangent. Finally, Fig.~\ref{fig:ViscNet_learning_curve}
shows the learning curve of the ViscNet neural network; no clear sign
of overfitting is present in this figure.

\subsection{Performance of the model \label{subsec:Metrics}}

There are many ways to assess the performance of predictive models;
metrics such as the coefficient of determination ($R^{2}$), the root
mean squared error (RMSE), the mean absolute error (MAE), and the
median absolute error (MedAE) give a holistic view on the performance
of regressors. Table \ref{tab:Experiments-metrics} shows these metrics
of ViscNet for the cross-validation experiment and the prediction
of the training, validation, and testing datasets. More information
on these metrics can be found in the \apdx{}. As expected, the training
dataset metrics are better than those of the validation dataset, which
in turn are better than those of the test dataset. These differences
reflect the influence that these datasets have in the training of
the model: the training dataset was used to change the weights and
bias of the network, the validation dataset was used to stop training
before it starts overfitting, and the test dataset had no influence
whatsoever in the training process. The cross-validation experiment
metrics are comparable to the metrics of the validation dataset, as
expected.

\begin{table}
\caption{Metrics of ViscNet for the cross-validation experiment and the training,
validation, and test datasets. The symbol $\uparrow$ indicates that
the higher the metric, the better, whereas the symbol $\downarrow$
indicates the opposite. The cross-validation column values are the
mean and standard deviation of the metrics for the 10 folds. \label{tab:Experiments-metrics} }

\centering{}%
\begin{tabular}{ccccc}
\toprule 
Metric & Training & Validation & Cross-val. & Test\tabularnewline
\midrule
\midrule 
$R^{2}$ ($\uparrow$) & 0.99 & 0.98 & \num{0.980(5)} & 0.97\tabularnewline
RMSE ($\downarrow$) & 0.58 & 0.88 & \num{0.9(1)} & 1.1\tabularnewline
MAE ($\downarrow$) & 0.42 & 0.64 & \num{0.60(8)} & 0.78\tabularnewline
MedAE ($\downarrow$) & 0.33 & 0.46 & \num{0.38(6)} & 0.53\tabularnewline
\bottomrule
\end{tabular}
\end{table}

The performance metrics obtained here were not as good as those reported
by Tandia et al.~\citep{tandia2019machine}. They achieved an impressive
$R^{2}$ value of 0.9999 and RMSE of 0.04 for the best architecture
on their validation dataset. Possible explanations for this difference
are related to the quality of the data, the number of chemical compounds
used for training, and the strategy to split the dataset. Tandia et
al.\ used a proprietary dataset owned by Corning Inc.\ that presumably
has much less variance than the dataset collected from SciGlass. The
number of chemical compounds used for training was 9 in Ref.~\citep{tandia2019machine}
and 39 in this work; it is more difficult for a model to generalize
in a diverse chemical domain. Finally, it is not stated if the dataset
spliting strategy used in Ref.~\citep{tandia2019machine} is similar
to the one used in this work (validation and test dataset are made
of liquids that are not present in the training dataset) or if it
is the commonly used random splitting; the expectation is that the
latter would yield better metrics as interpolation is easier than
extrapolation.

\begin{figure}
\begin{centering}
\includegraphics[width=0.8\columnwidth]{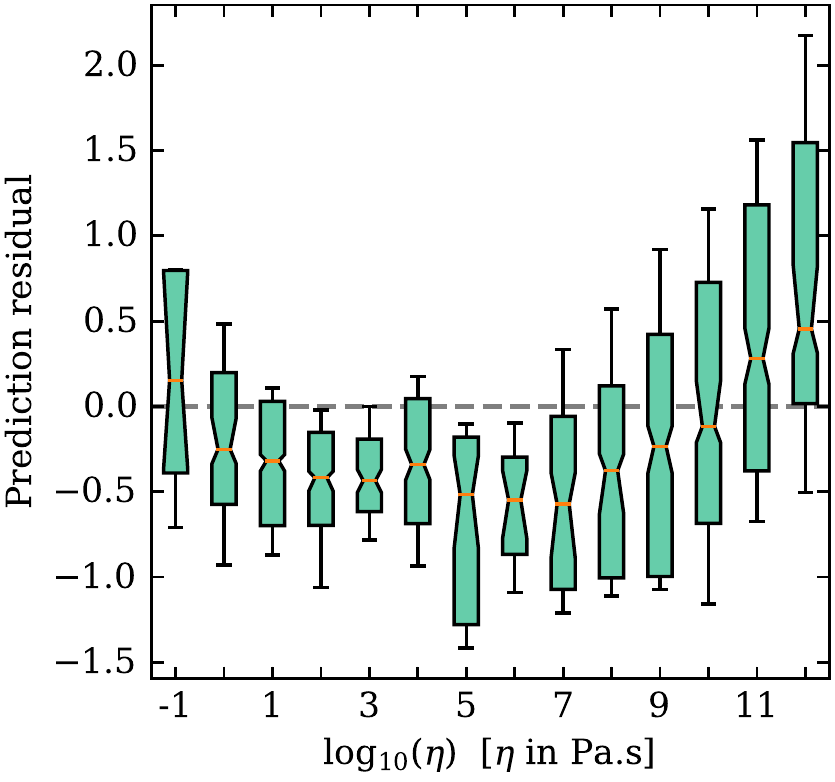}
\par\end{centering}
\caption{Boxplot of the prediction residual versus the reported value of viscosity
for the test dataset. The boxes are bound by \SI{25}{\percent} and
\SI{75}{\percent} percentiles, and the error caps represent \SI{67}{\percent}
of the data. The median is shown as a horizontal orange line, and
the notch of the median represents its \SI{95}{\percent} confidence
interval. \label{fig:Boxplot-of-the}}
\end{figure}

\begin{figure*}[!t]
\begin{centering}
\includegraphics[width=0.7\textwidth]{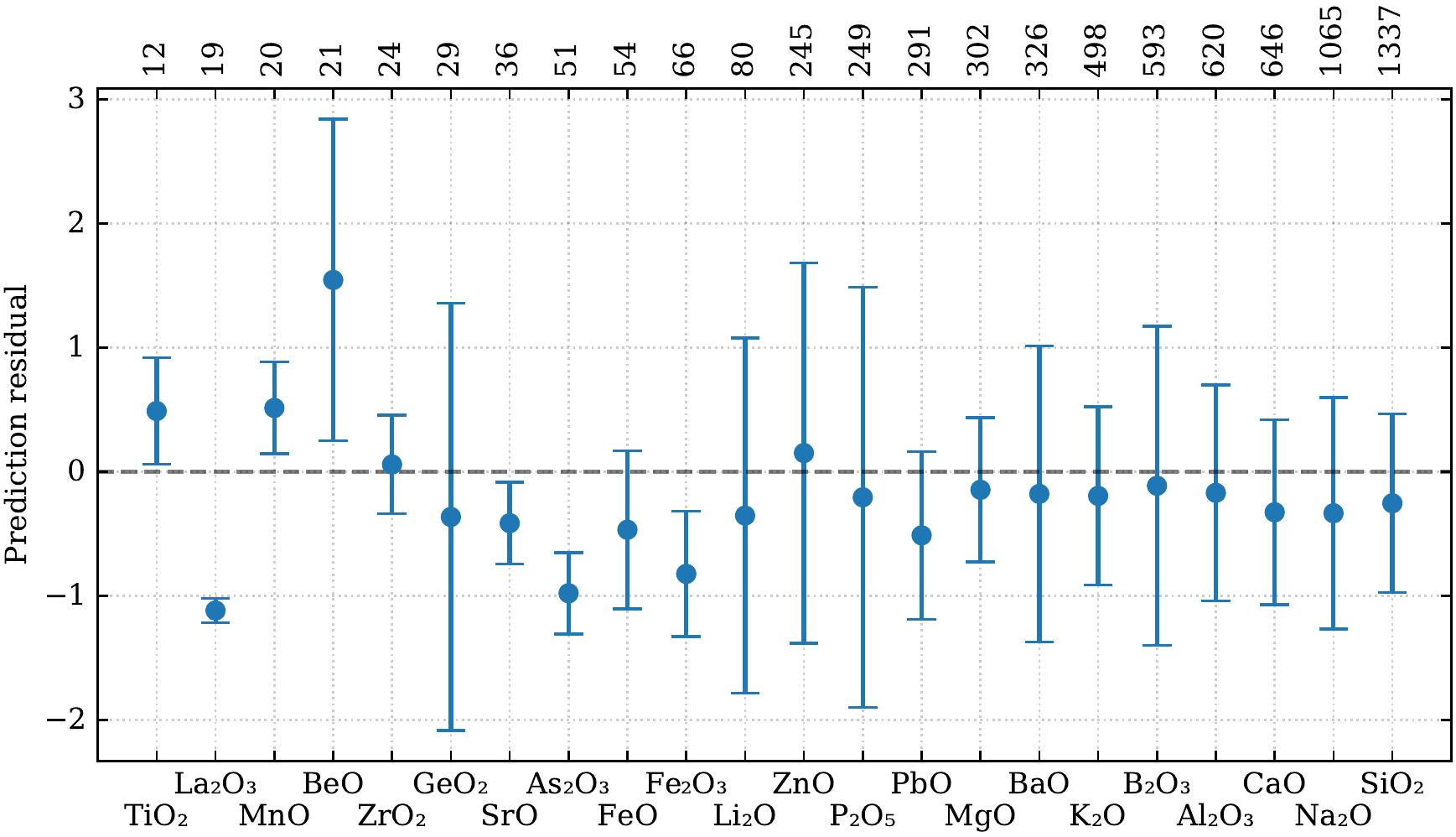}
\par\end{centering}
\caption{Mean and standard deviation of the prediction residual versus the
chemical compound for the test dataset. The top $x$-axis shows the
number of data points in the test dataset that have the respective
compound; it is organized in crescent order from left to right. \label{fig:Mean-and-standard}}
\end{figure*}

Another strategy to assess the performance of ViscNet is by looking
into the prediction residuals. Figures \ref{fig:Boxplot-of-the} and
\ref{fig:Mean-and-standard} show the prediction residual versus the
viscosity range and the chemical compound in the liquid, respectively.
Both plots show only predictions for the test dataset, as these predictions
suggest how well the model can predict data that it has never seen,
thus helping assess the generalization capabilities of the model.

Figure \ref{fig:Boxplot-of-the} shows that the prediction's uncertainty
is\linebreak{}
higher for higher values of viscosity. However, the median prediction
residual is in the range of \num{-0.5} and 0.5, which is expected
as the MedAE for the test dataset is about 0.5. Figure \ref{fig:Mean-and-standard}
shows that some compounds such as \ce{GeO2}, \ce{LiO2}, \ce{ZnO},
\ce{P2O5}, \ce{B2O3}, \ce{BeO}, and \ce{BaO} have a standard deviation
of the prediction residual greater than one. Care should be taken
when using ViscNet to predict liquids having these compounds. 

\begin{figure}[!t]
\begin{centering}
\includegraphics[width=0.8\columnwidth]{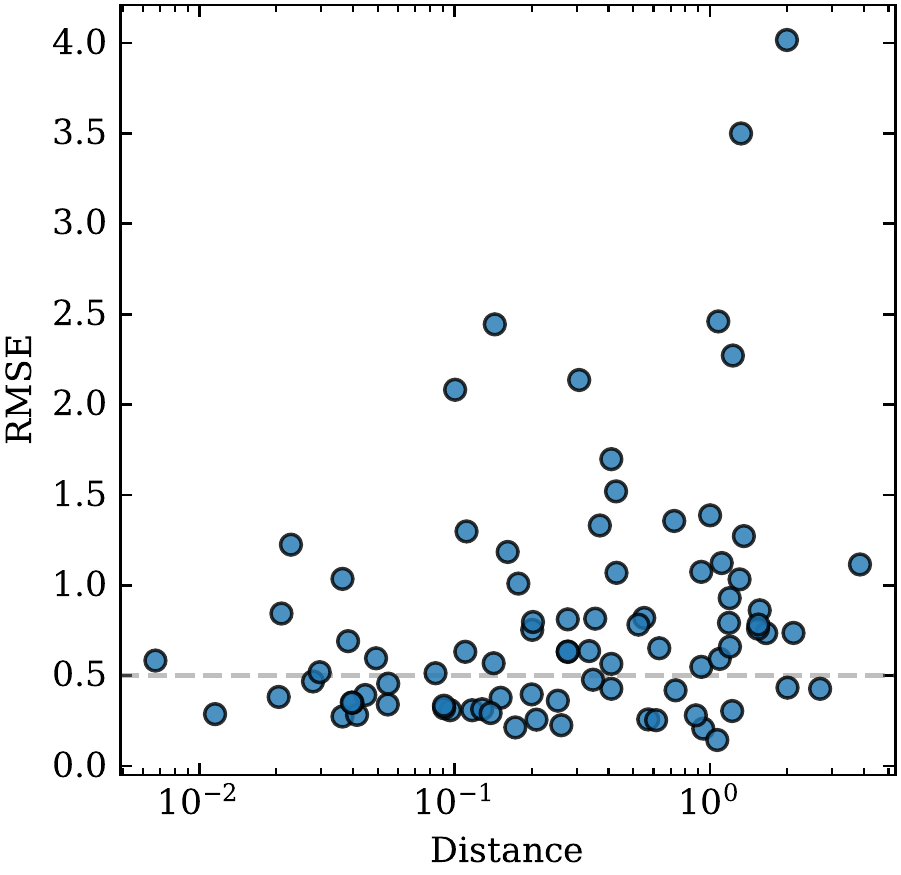}
\par\end{centering}
\caption{RMSE of prediction of the liquids in the test dataset versus their
distance from the training and validation domain (see text). Each
point represents a different liquid. The dashed gray line shows RMSE
= 0.5. \label{fig:RMSE_distance}}
\end{figure}

All the liquids in the test dataset have compositions that were not
present in the datasets used for training and validating the model.
This splitting strategy was a design choice to promote better extrapolation
instead of better interpolation of viscosity. A question that arises
is if the prediction accuracy is related to how distant the composition
is to the domain of training. The distance used was the Canberra distance
between the composition vector of the liquid and its closest neighbor
in the training and validation domain. The Canberra distance is a
weighed $L_{1}$ distance; $L_{2}$ distances (such as Euclidean)
are not recommended for problems with more than three dimensions \citep{aggarwal2001surprising}.
Figure \ref{fig:RMSE_distance} shows that most liquids in the test
dataset have an RMSE close to 0.5, but as the distance increases,
so do the chances of having a higher value of RMSE. 

\begin{figure}[t]
\begin{centering}
\includegraphics[width=1\columnwidth]{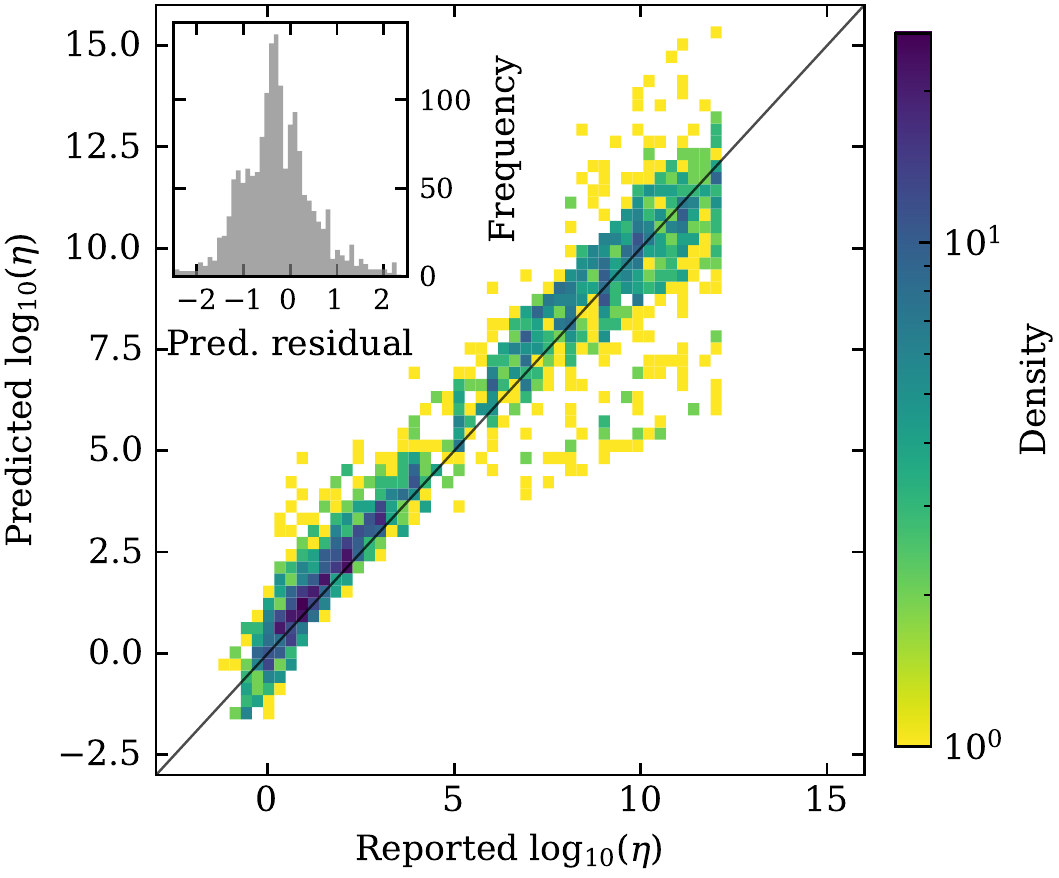}
\par\end{centering}
\caption{2D histogram of predicted versus reported values of $\log_{10}\left(\eta\right)$
for the test dataset. Each square has a corner of 0.3, and the continuous
black line is the identity line. The inset is the histogram of the
prediction residuals. \label{fig:correlation}}
\end{figure}

Figure \ref{fig:correlation} shows a 2D histogram with the correlation
between the predicted and reported viscosity values of the test dataset.
Most of the data points are close to the identity line, but a noticeable
spread is present, especially in the region of higher viscosity (as
already suggested by Fig.~\ref{fig:Boxplot-of-the}). The inset of
this plot shows a histogram of the prediction residuals. The model
has a small bias towards predicting higher values of viscosity than
those reported.

\begin{figure}[!t]
\begin{centering}
\includegraphics[width=0.8\columnwidth]{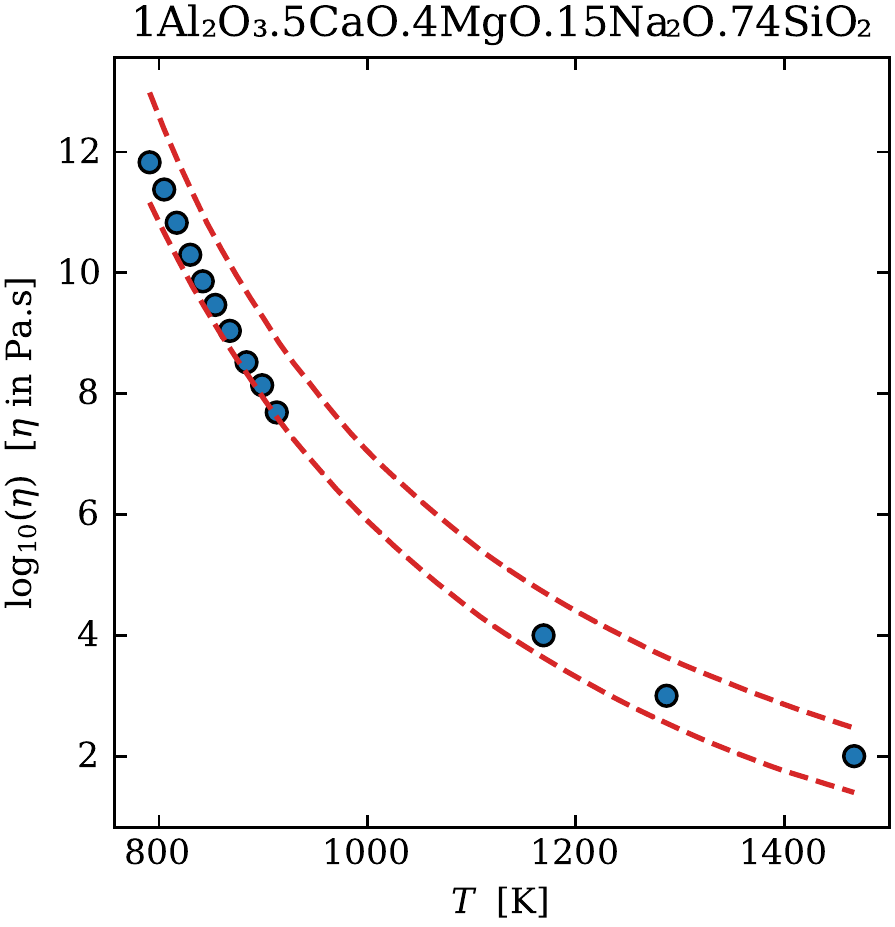}
\par\end{centering}
\caption{Base-10 logarithm of viscosity versus temperature for \ce{Al2O3.5CaO.4MgO.15Na2O.74SiO2},
one of the liquids in the test dataset. The blue circles are experimental
data, and the dashed red lines are the model uncertainty bands with
a confidence interval of \SI{95}{\percent}. \label{fig:individual_liquid_example}}
\end{figure}

\begin{figure}[!t]
\begin{centering}
\includegraphics[width=0.8\columnwidth]{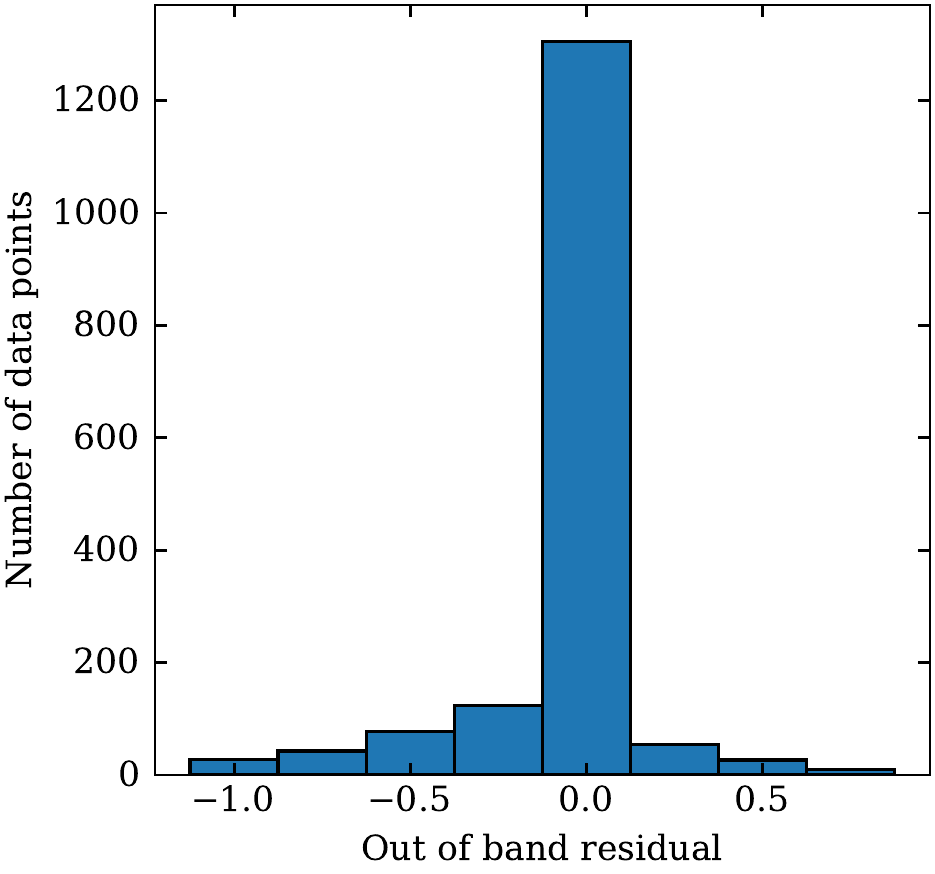}
\par\end{centering}
\caption{Histogram of out of band residual values of the test dataset prediction.
The out of band residual is zero if the data point is within the prediction
bands or the difference between the reported viscosity value and the
closest uncertainty band otherwise (in base-10 logarithm scale). \label{fig:out-of-band-residual}}
\end{figure}

The final analysis of the ViscNet performance consists of looking
at the data points and the model prediction for all the liquids in
the test dataset. Figure \ref{fig:individual_liquid_example} shows
this analysis for one of the liquids; individual plots for all the
other 84 liquids are shown in the \apdx{}. The uncertainty of prediction
represents a confidence interval of \SI{95}{\percent} and was computed
by Monte Carlo dropout \citep{gal2016dropout} with \num{1000} random
samples. For many liquids in the test dataset, the uncertainty bands
contain the experimental data or predict the general trend of viscosity
correctly. Figure \ref{fig:out-of-band-residual} shows that \num{1200}
data points of the test dataset (about \SI{70}{\percent}) are within
the prediction bands of the model. There are problematic compositions,
as expected, where neither the magnitude nor tem\-per\-a\-ture-dependence
of viscosity was adequately predicted. As already discussed, predicting
the viscosity for a liquid too far from the training and validation
domain increases the chances of a wrong prediction. A (labor intense)
solution is to collect more data to expand the training domain; another
solution is to train the model with structural features in addition
to chemical features.

\subsection{Parameters of viscosity}

\begin{figure*}[!t]
\begin{centering}
\subfloat[]{\begin{centering}
\includegraphics[width=0.45\textwidth]{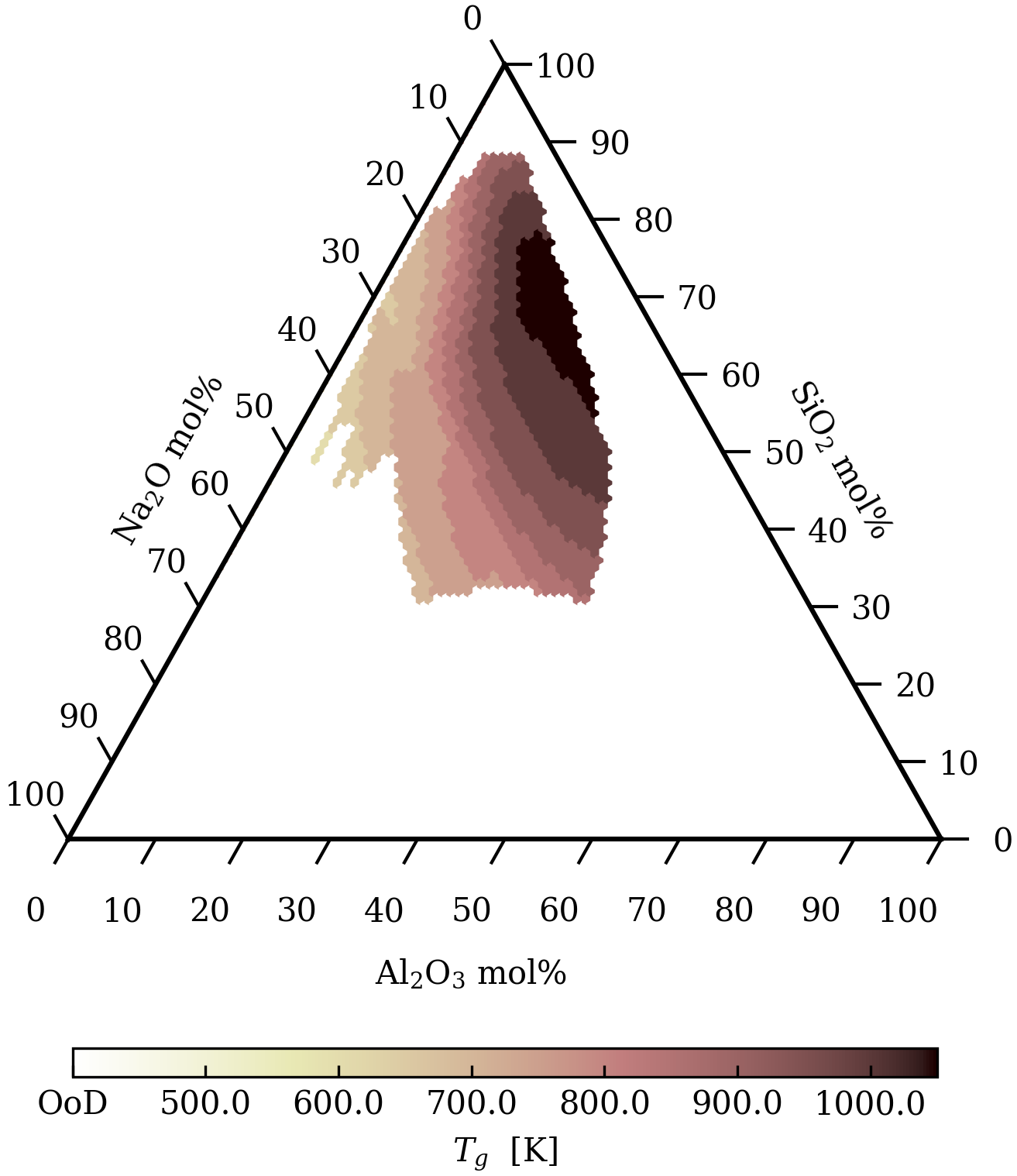}
\par\end{centering}
}\subfloat[]{\begin{centering}
\includegraphics[width=0.45\textwidth]{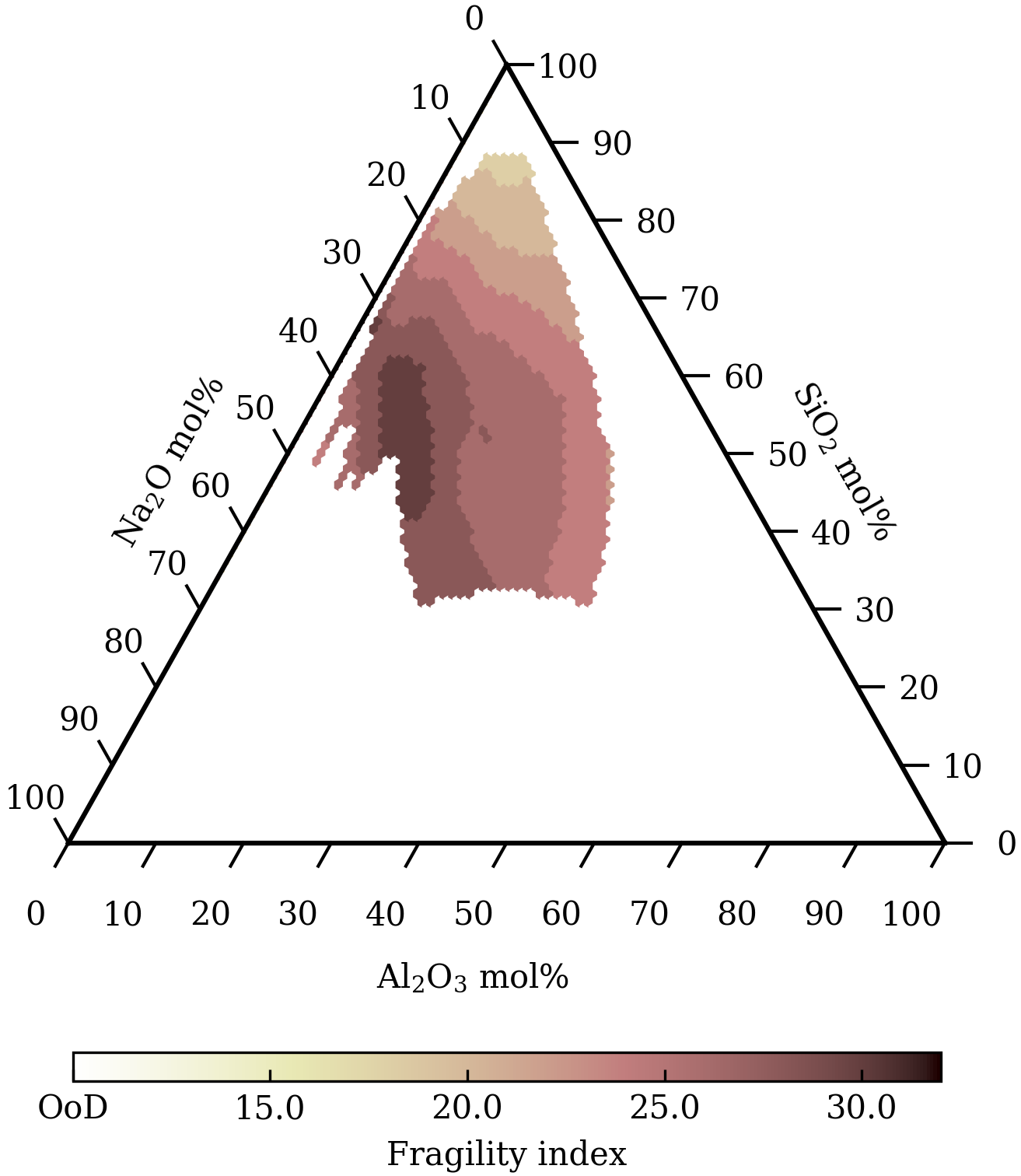}
\par\end{centering}
}
\par\end{centering}
\caption{Ternary plot of (\textbf{a}) $T_{g}$ and (\textbf{b}) fragility index
for the system \ce{SiO2-Na2O-Al2O3} predicted by ViscNet. For better
visualization, $T_{g}$ values were rounded to the closest multiple
of 50 and fragility index values to the closest multiple of 2. Only
data with a distance of 0.5 or less of the closest neighbor in the
training and validation domain are shown in the plots (see Fig.~\ref{fig:RMSE_distance}
and related discussion). Out of domain (OoD) region is shown in white.
\label{fig:Fragility-index-ternary}}
\end{figure*}

As already mentioned, one advantage of the gray-box approach, in contrast
with the black-box approach, is the direct access to the parameters
of the viscosity model. Figure \ref{fig:Fragility-index-ternary}
leverages this advantage by showing the prediction of the glass transition
temperature and the fragility index for the ternary system \ce{SiO2-Na2O-Al2O3},
one of the base systems currently used for developing scratch-resistant
display screens. The predictions of both properties are contained
only in the region that the composition vector has a distance of 0.5
or less to its closest neighbor in the training and validation domain.
As shown in the previous section, going too far away from this domain
increases the chance of a wrong prediction.

The plots in Fig.~\ref{fig:Fragility-index-ternary} show that the
glass transition temperature decreases significantly with the addition
of \ce{Na2O} but changes at a much lower rate with the addition of
\ce{Al2O3}. While this is well known in the glass community (\ce{Na2O}
is a modifier and \ce{Al2O3} is an intermediate compound), it is
interesting to see that the model could capture this behavior from
data. The fragility index follows a similar trend, but with $m$ increasing
instead of decreasing with the addition of \ce{Na2O} and \ce{Al2O3}.

\subsection{Transfer learning}

As already mentioned, some hyperparameters were fixed from the very
beginning, during the machine learning pipeline design. The viscosity
model and the backpropagation loss function are two examples.

Another three-parameter viscosity model is the VFT empirical equation
(Eq.~\eqref{eq:VFT}) \citep{vogel1921temperatureabhangigketsgesetz,fulcher1925analysis,tammann1926abhangigkeit},
which has historical significance and is still often used in scientific
research. It is possible to build a machine learning pipeline similar
to that shown in Fig.~\ref{fig:Schematics-pipeline}, but with the
VFT equation instead of the MYEGA equation, and use the weights and
bias of ViscNet as a starting point for the new model; this process
is called \emph{transfer learning}. The resulting model following
this strategy will be called ViscNet-VFT. 

\begin{multline}
\log_{10}\left(\eta\left(T,\eta_{\infty},T_{g},m\right)\right)=\log_{10}\left(\eta_{\infty}\right)\\
+\frac{\left(12-\log_{10}\left(\eta_{\infty}\right)\right)^{2}}{m\left(T/T_{g}-1\right)+\left(12-\log_{10}\left(\eta_{\infty}\right)\right)}\label{eq:VFT}
\end{multline}

The same procedure can be used to explore different backpropagation
loss functions. The loss function of the original pipeline was the
MSE, which is sensitive to outliers. By using transfer learning, we
can change this loss function to one that is robust against outliers,
such as the Huber loss \citep{huber1964robust}. The resulting model
following this strategy will be called ViscNet-Huber.

One advantage of using transfer learning is that it is significantly
faster than developing a model bottom-up. The HP tuning routine for
ViscNet took about a day and a half of computing time while training
both new models using transfer learning took only a few minutes.

The prediction power of ViscNet-VFT is comparable to that of ViscNet.
This result is expected as the MYEGA and VFT equations both yield
similar results in the temperature range where experimental data is
available. They significantly differ, however, if the model is extrapolated
to regions where the viscosity is higher than \SI{e12}{Pa.s}.

Any other three-parameter viscosity model that can be formulated in
function of $\eta_{\infty}$, $T_{g}$, and $m$ (such as the AM equation
\citep{avramov1988effect}, for example) can also be used to create
other models via transfer learning, similarly to what was performed
with the VFT equation. However, the expectation is that no significant
differences in prediction power will be observed in the temperature
range where experimental data are available, as was the case with
VFT.

\begin{figure*}[!t]
\begin{centering}
\subfloat[\label{fig:ViscNet-Huber_boxplot}]{\begin{centering}
\includegraphics[width=0.45\textwidth]{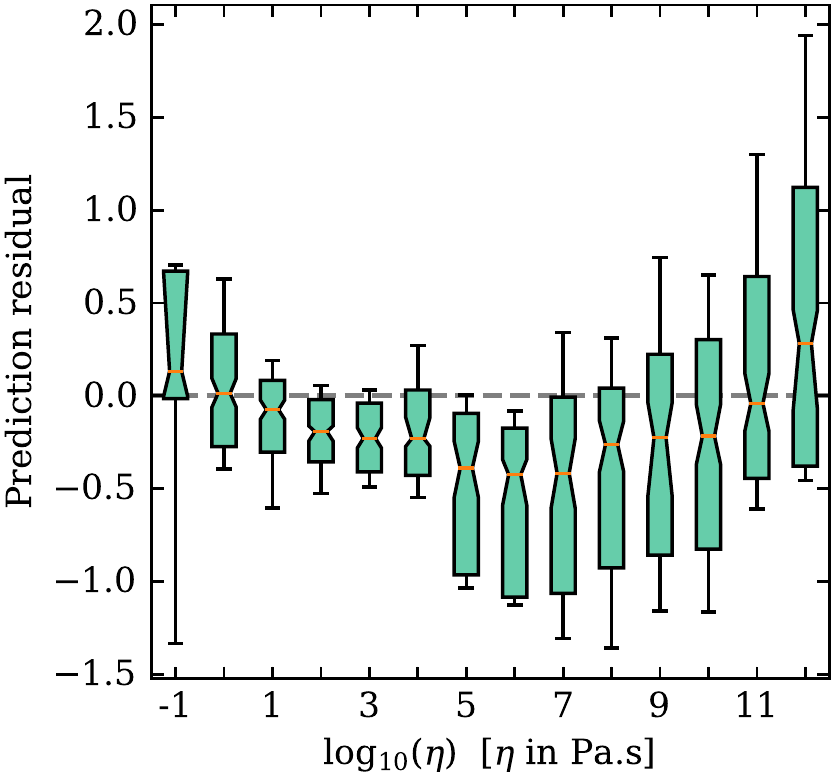}
\par\end{centering}
}\subfloat[]{\begin{centering}
\includegraphics[width=0.45\textwidth]{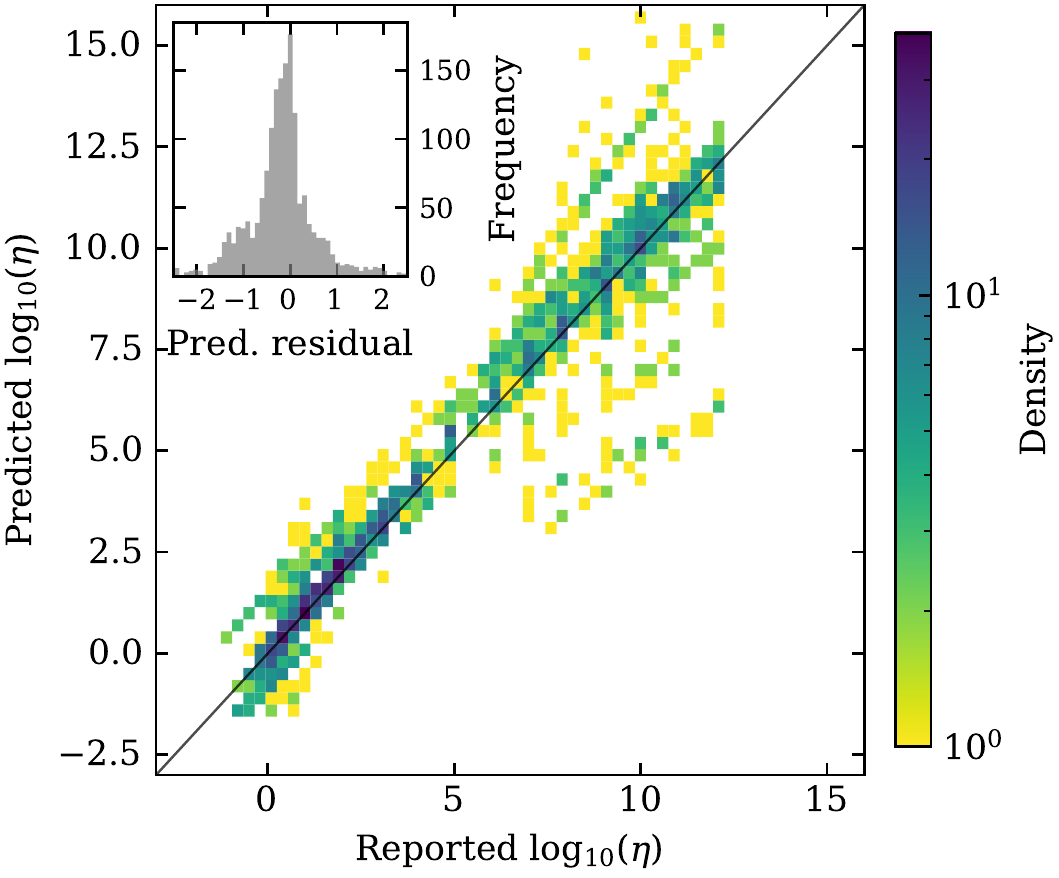}
\par\end{centering}
}
\par\end{centering}
\begin{centering}
\subfloat[]{\begin{centering}
\includegraphics[width=0.7\textwidth]{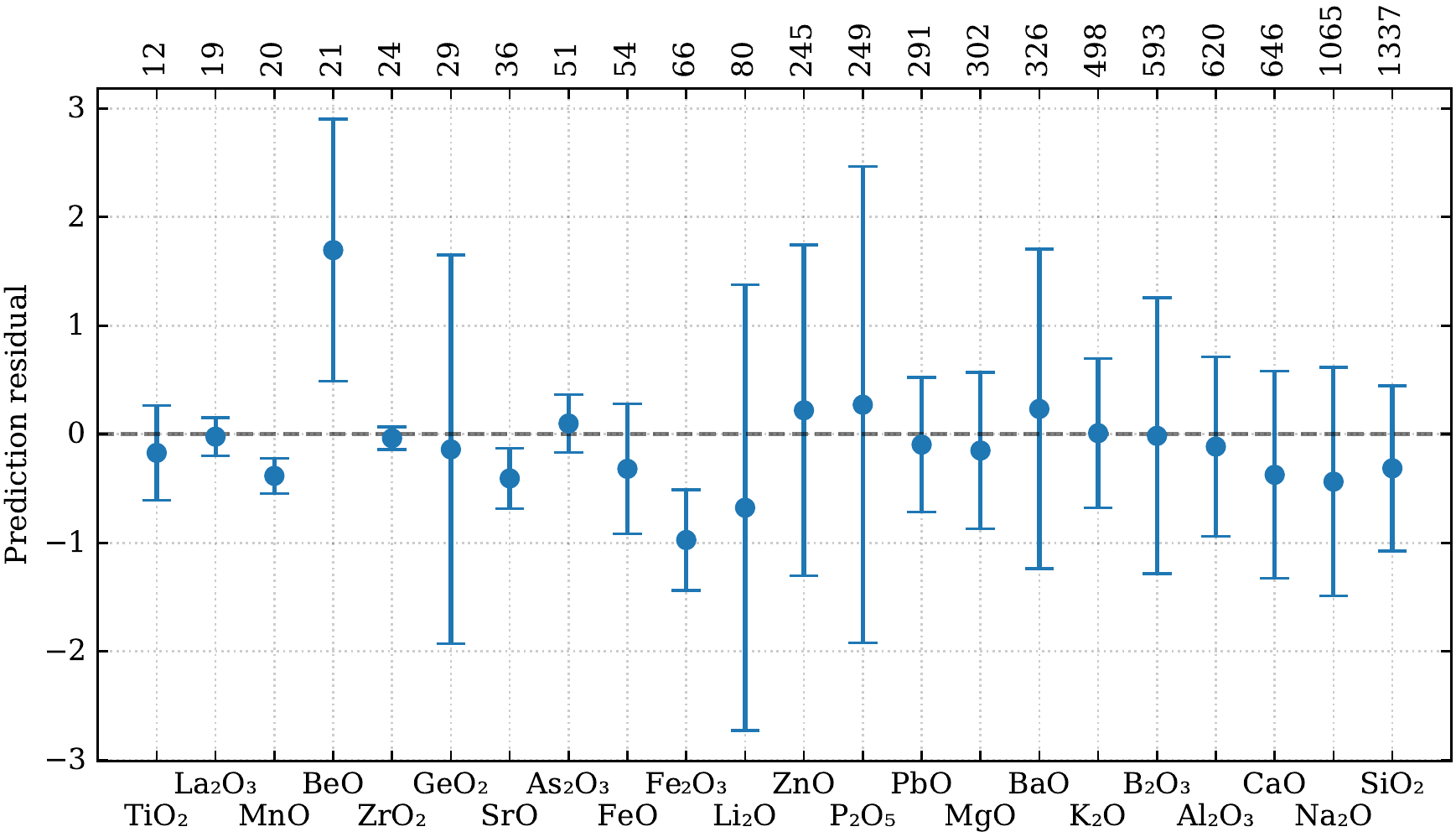}
\par\end{centering}
}
\par\end{centering}
\caption{Results of the ViscNet-Huber model for the test dataset: (\textbf{a})
boxplot of the prediction residual versus the reported value of viscosity;
(\textbf{b}) 2D histogram of predicted versus reported values of $\log_{10}\left(\eta\right)$;
and (\textbf{c}) mean and standard deviation of the prediction residual
versus the chemical compound. See Section \ref{subsec:Metrics} for
information on how these plots were made. \label{fig:ViscNet-Huber}}
\end{figure*}

Using transfer learning to change the loss function had an interesting
consequence: ViscNet-Huber is better at predicting high-temperature
viscosity than ViscNet (see Fig.~\ref{fig:ViscNet-Huber_boxplot}).
This temperature region is particularly important for processing glasses
via melt and quench, which is the most used route to process commercial
glasses.

Figure \ref{fig:ViscNet-Huber} shows some results on the prediction
power of ViscNet-Huber. Additional plots are available in the \apdx{}
for the interested reader. Apparently, this work is the first to use
transfer learning in the context of glass-forming liquids.

\subsection{Reproducibility and data availability \label{subsec:Availability-and-reproducibility}}

This work was entirely performed using open-source software and the
SciGlass database, which has a permissive license. Code containing
all the necessary functions to load the data and train the machine
learning pipelines discussed here is publicly available on GitHub
(\url{https://github.com/drcassar/viscnet}) and Zenodo \citep{cassar_viscnet},
licensed as free software under the GPL3. This code leverages deterministic
routines for training the NNs provided by the \texttt{PyTorch-Lightning}
module; thus interested readers can reproduce the exact models reported
here. Pretrained networks for ViscNet, ViscNet-Huber, and ViscNet-VFT
are also available in the repository.

Due to the free and open-source nature of the data and the code, anyone
can extend the procedures presented here to better meet their needs,
for example, including new features for training the models or training
the models with different datasets.

\section{Conclusion}

This work aimed to build a machine learning pipeline to predict the
tem\-per\-a\-ture-dependence of the viscosity of oxide liquids,
inspired by a recent gray-box neural network developed by Tandia et
al.~\citep{tandia2019machine}, who embedded a physical model in
the pipeline. This work introduced a pre-processing unit with a chemical
feature extractor, which changes the feature domain from chemical
composition to chemical properties. 

The predictive model was focused on extrapolation, and it was able
to predict the viscosity of the 85 liquids in the test dataset with
an $R^{2}$ of 0.97. About \SI{70}{\percent} of the data points in
the test dataset where within the uncertainty bands of the model's
prediction. However, the chances of a wrong prediction increases with
the distance to the closest neighbor in the training and validation
datasets. 

The performance and speed of the predictive models can be exploited
to guide the development of new glasses. The viscosity prediction
can help in selecting compositions with a particular viscosity behavior
or determining process variables. The fragility index and glass transition
temperature predictions can help in selecting compositions with desired
properties for specific applications.

All code used in this work was built with reproducibility in mind,
using open-source Python modules. Both data and code are available
for anyone interested, at no cost, and with a permissive license:
the hope is that this free and open framework for property prediction
could be used and improved by the community to accelerate the development
of new materials.

\section*{Acknowledgments}

The author is thankful for the São Paulo State Research Foundation
support (FAPESP grant number 2017/\LyXZeroWidthSpace 12491-0) as well
as for the Nippon Sheet Glass Foundation overseas research grant.
The author also thanks John Mauro, Adama Tandia, Bruno Rodrigues,
and Collin \linebreak{}
Wilkinson for insightful comments and suggestions; and Carolina Zanelli
for text revision.

\section*{Data availability statement}

The viscosity data used in this work comes from the SciGlass database.
This database is available under an ODC Open Database License (ODbL)
at \url{https://github.com/epam/SciGlass}. See Section \ref{subsec:Availability-and-reproducibility}
for more information on reproducing this research.

\section*{Competing interest statement}

The author declares no competing financial or non-financial interests. 

\bibliographystyle{elsarticle-num}
\bibliography{library}

\appendix
\begin{center}
{\huge{}Appendix }{\huge\par}
\par\end{center}

\section{z-score}

In the pre-processing unit of the machine learning pipe\-line, a
normalization step computes the z-score $z_{i}$ of each feature that
will be fed to the NN. This process (Eq.~\eqref{eq:z-score}) is
performed for each feature $f_{i}$ by subtracting the mean value
of this feature ($\mu$) and scaling the data to unit variance by
dividing by the standard deviation of this feature ($s_{d}$).

\begin{equation}
z_{i}=\frac{f_{i}-\mu}{s_{d}}\label{eq:z-score}
\end{equation}

\section{Evaluation metrics}

Four metrics were computed in Section \ref{subsec:Metrics} and are
discussed here.

The coefficient of determination, $R^{2}$, has various definitions.
Here it is used to test the relationship between the predicted and
the reported base-10 logarithm of viscosity ($\hat{y}$ and $y$,
respectively). The ideal relationship is a linear model with no intercept,
for which the $R^{2}$ can be computed via Eq.~\eqref{eq:Rsquared}.
The value of $R^{2}$ is dimensionless and between zero and one, indicating,
respectively, no correlation and a perfect correlation between predicted
and reported viscosity values.

\begin{equation}
R^{2}=1-\frac{\sum_{i}^{n}\left(y_{i}-\hat{y_{i}}\right)^{2}}{\sum_{i}^{n}y_{i}^{2}}\label{eq:Rsquared}
\end{equation}

The root mean square error, RMSE, is a measure of the difference between
$y$ and $\hat{y}$. It is the square root of the mean square error,
as can be seen in Eq.~\eqref{eq:RMSE}, and it has the advantage
of being in the same unit as $y$. The lower the RMSE, the better.

\begin{equation}
\mathrm{RMSE}=\sqrt{\frac{1}{n}\sum_{i}^{n}\left(y_{i}-\hat{y}_{i}\right)^{2}}\label{eq:RMSE}
\end{equation}

The mean absolute error MAE is the average of the absolute errors.
It is also a measure of the difference between $y$ and $\hat{y}$,
but differently from RMSE, each error contributes equally, and the
residuals are not squared. This metric has the same unit as $y$ and
is computed using Eq.~\eqref{eq:MAE}. The lower the MAE, the better.

\begin{equation}
\mathrm{MAE}=\frac{\sum_{i}^{n}\left|y_{i}-\hat{y}_{i}\right|}{n}\label{eq:MAE}
\end{equation}

The median absolute error MedAE is similar to MAE, but instead of
computing the average residual value, it computes the median value.
This metric is robust against outliers; it has the same unit as $y$
and is computed using Eq.~\eqref{eq:MedAE}. The lower the MedAE,
the better.

\begin{equation}
\mathrm{MedAE}=\mathrm{median}\left(\left|y_{1}-\hat{y}_{1}\right|,\left|y_{2}-\hat{y}_{2}\right|,\ldots,\left|y_{n}-\hat{y}_{n}\right|\right)\label{eq:MedAE}
\end{equation}

\section{Supplementary plots}

\begin{figure*}
\begin{centering}
\includegraphics[width=1\textwidth]{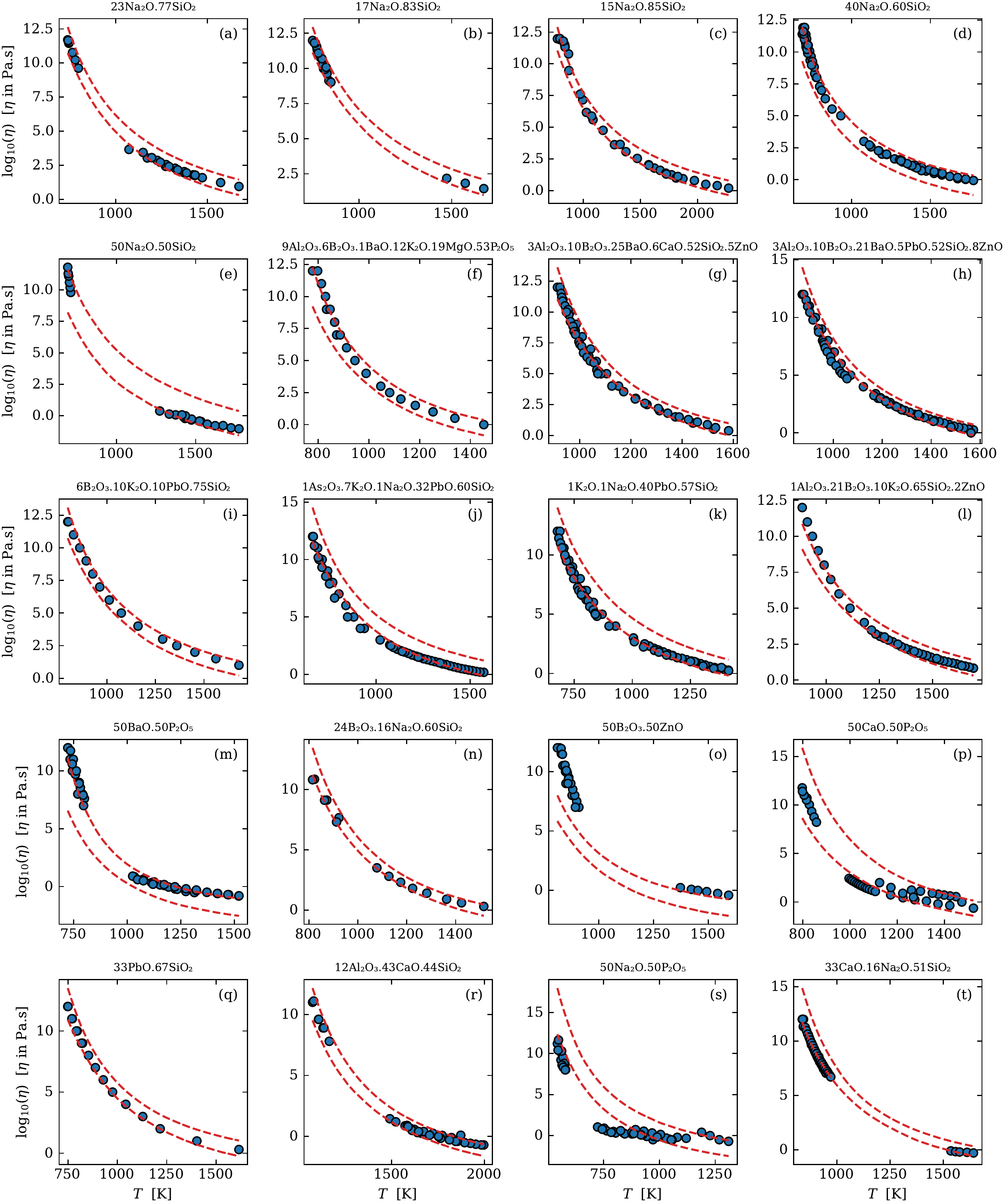}
\par\end{centering}
\caption{Base-10 logarithm of viscosity versus temperature for 20 liquids in
the test dataset. The blue circles are experimental data, and the
dashed red lines are the ViscNet prediction bands with a confidence
of \SI{95}{\percent}. \label{fig:1st_apdx_plot}}
\end{figure*}

\begin{figure*}
\begin{centering}
\includegraphics[width=1\textwidth]{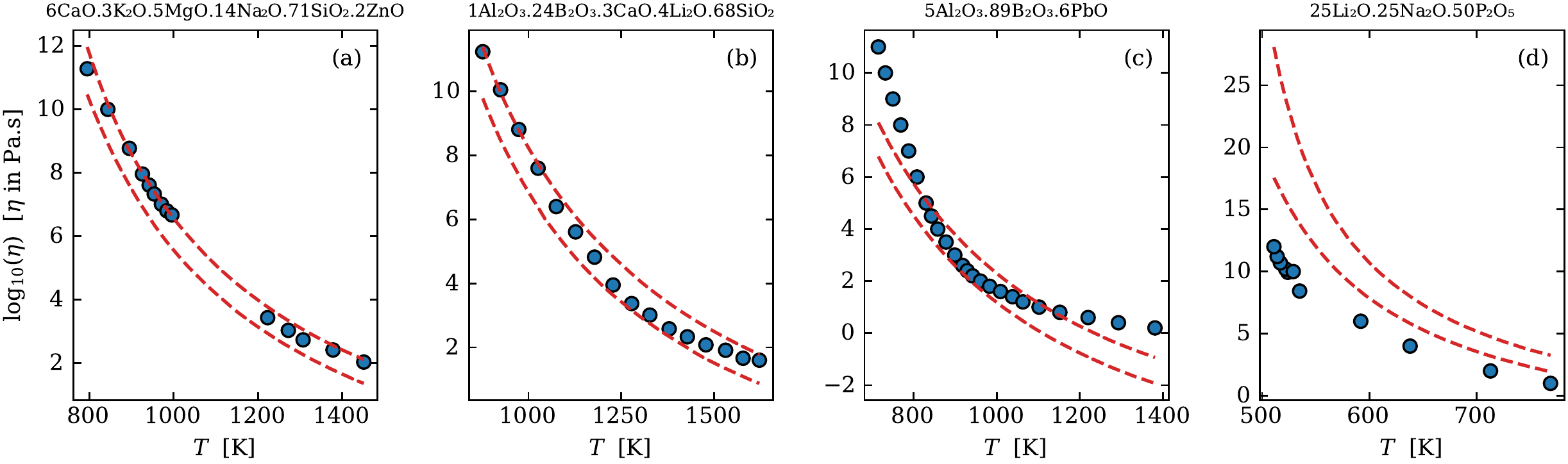}
\par\end{centering}
\caption{Base-10 logarithm of viscosity versus temperature for 4 liquids in
the test dataset. The blue circles are experimental data, and the
dashed red lines are the ViscNet-VFT prediction bands with a confidence
of \SI{95}{\percent}. \label{fig:Last_apdx_plot}}
\end{figure*}

Figures \ref{fig:1st_apdx_plot} to \ref{fig:Last_apdx_plot} show
plots on the performance of ViscNet, ViscNet-Huber, and ViscNet-VFT.

\begin{figure*}
\begin{centering}
\includegraphics[width=1\textwidth]{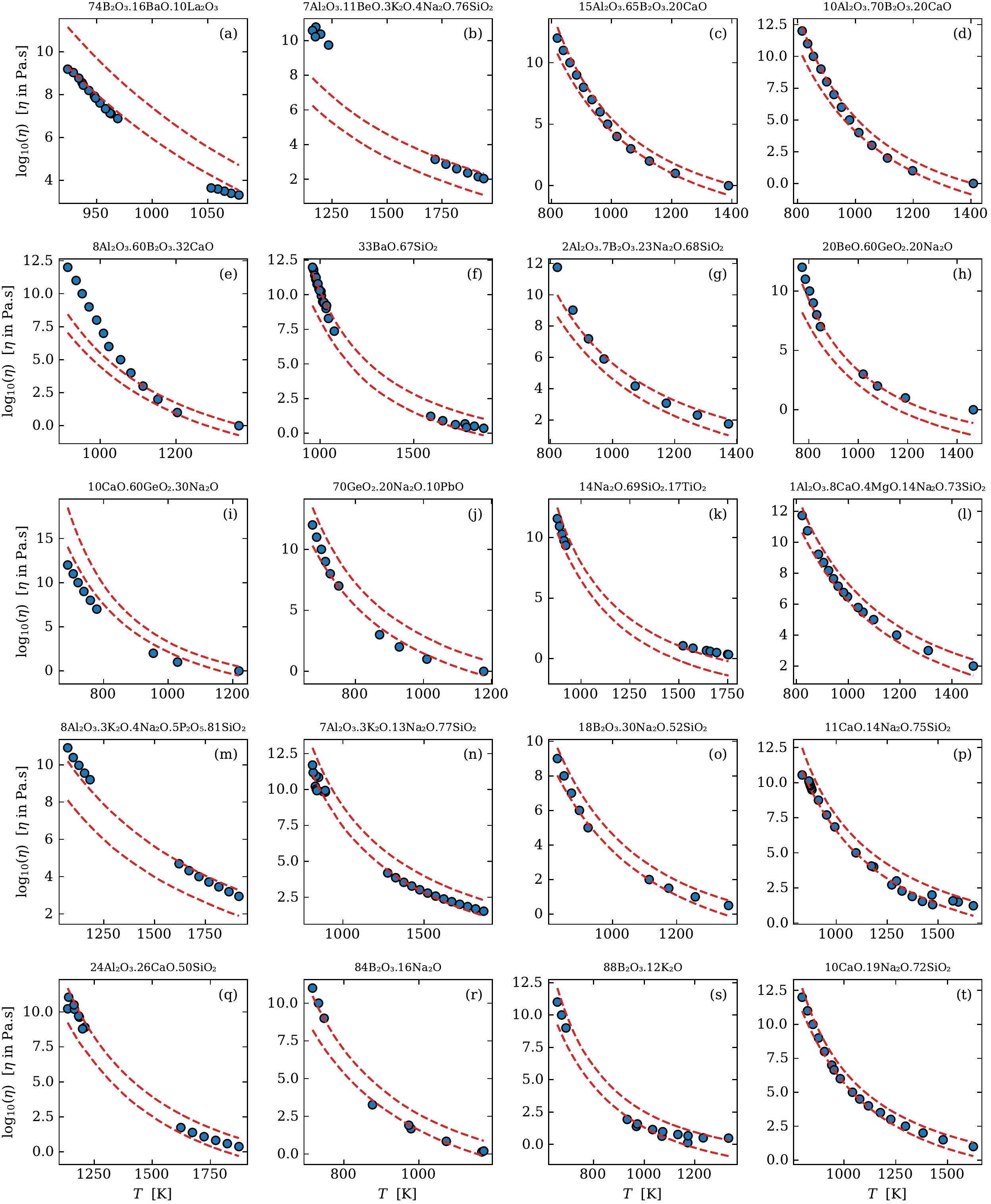}
\par\end{centering}
\caption{Base-10 logarithm of viscosity versus temperature for 20 liquids in
the test dataset. The blue circles are experimental data, and the
dashed red lines are the ViscNet prediction bands with a confidence
of \SI{95}{\percent}.}
\end{figure*}

\begin{figure*}
\begin{centering}
\includegraphics[width=1\textwidth]{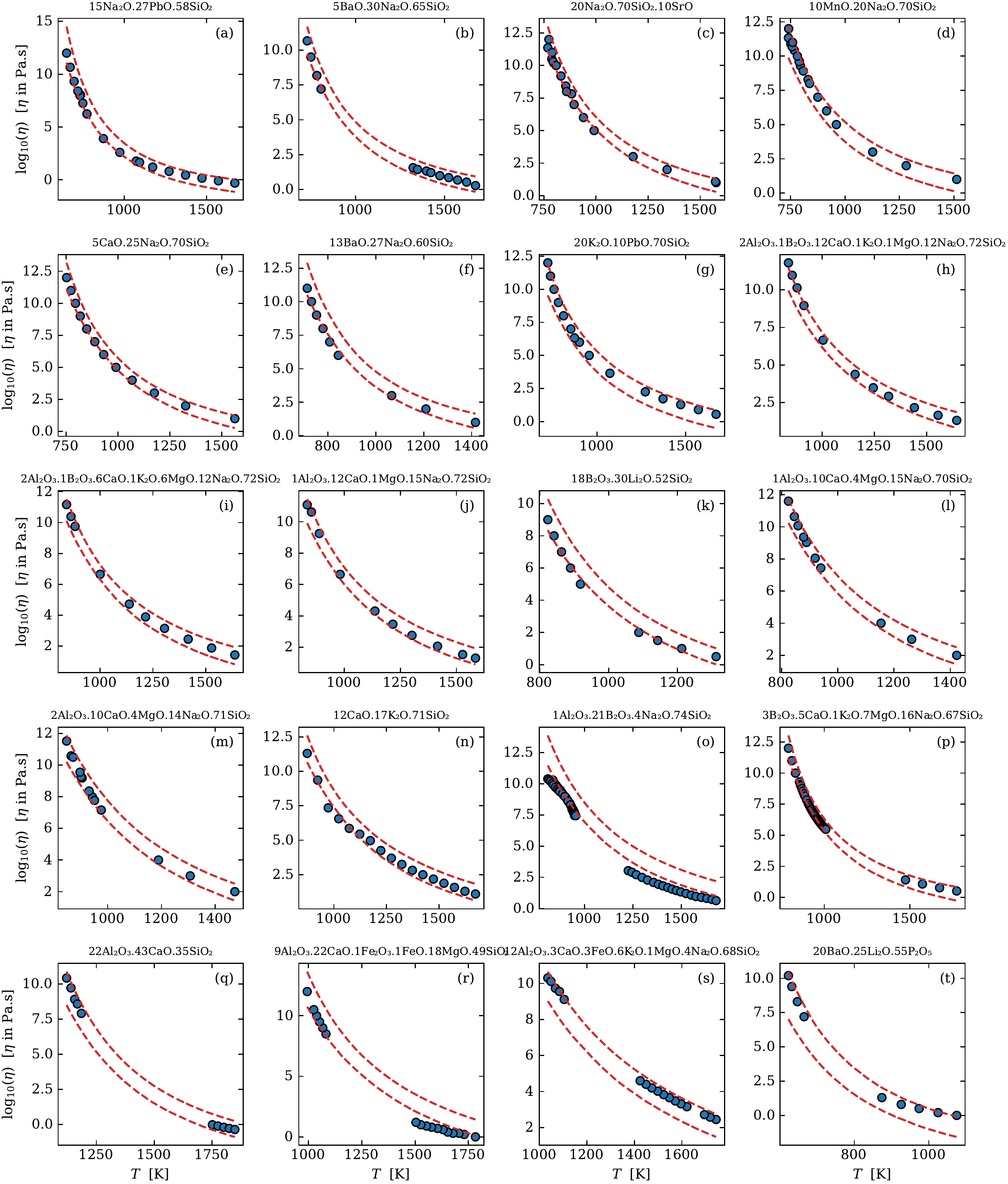}
\par\end{centering}
\caption{Base-10 logarithm of viscosity versus temperature for 20 liquids in
the test dataset. The blue circles are experimental data, and the
dashed red lines are the ViscNet prediction bands with a confidence
of \SI{95}{\percent}.}
\end{figure*}

\begin{figure*}
\begin{centering}
\includegraphics[width=1\textwidth]{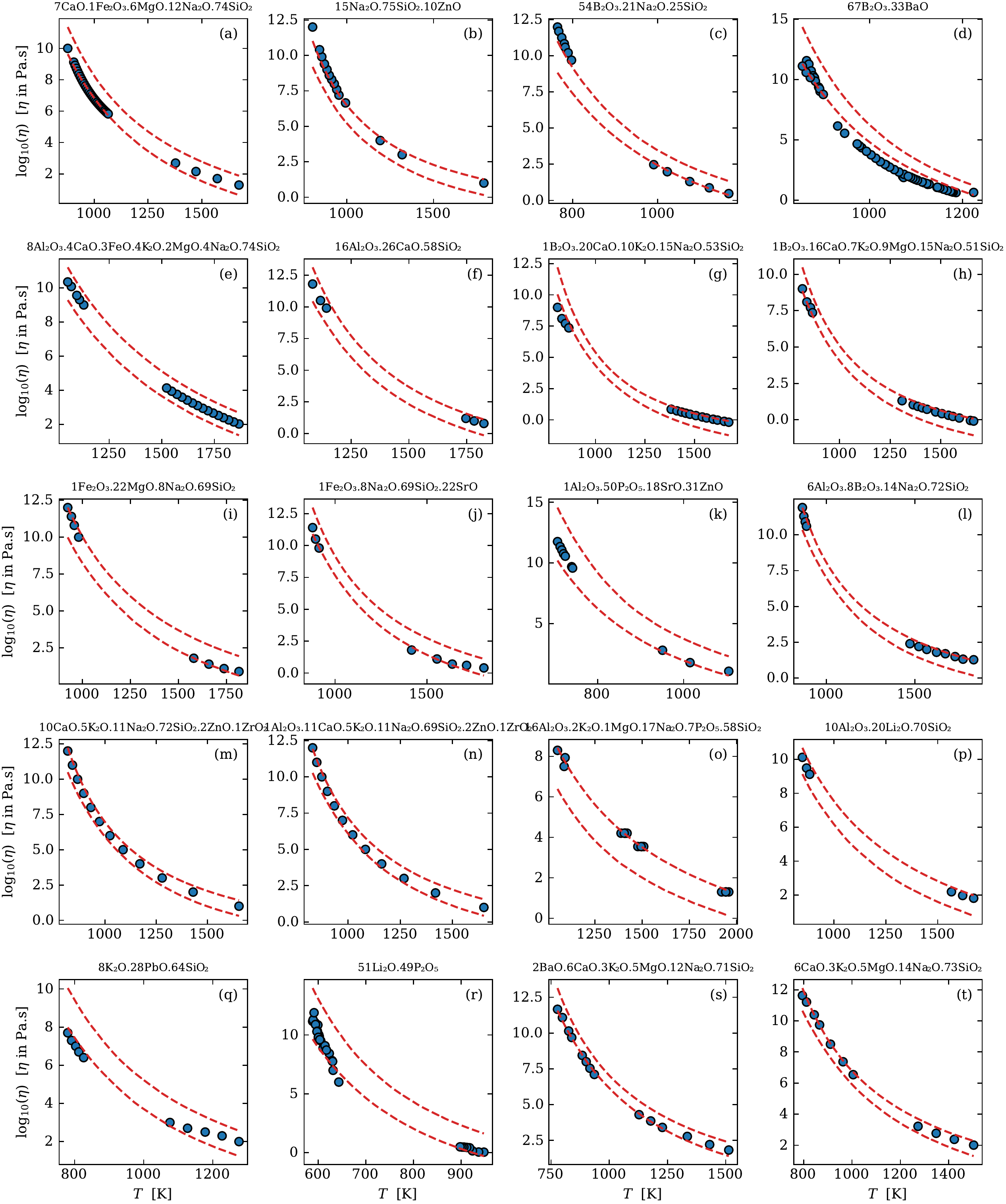}
\par\end{centering}
\caption{Base-10 logarithm of viscosity versus temperature for 20 liquids in
the test dataset. The blue circles are experimental data, and the
dashed red lines are the ViscNet prediction bands with a confidence
of \SI{95}{\percent}.}
\end{figure*}

\begin{figure*}
\begin{centering}
\includegraphics[width=1\textwidth]{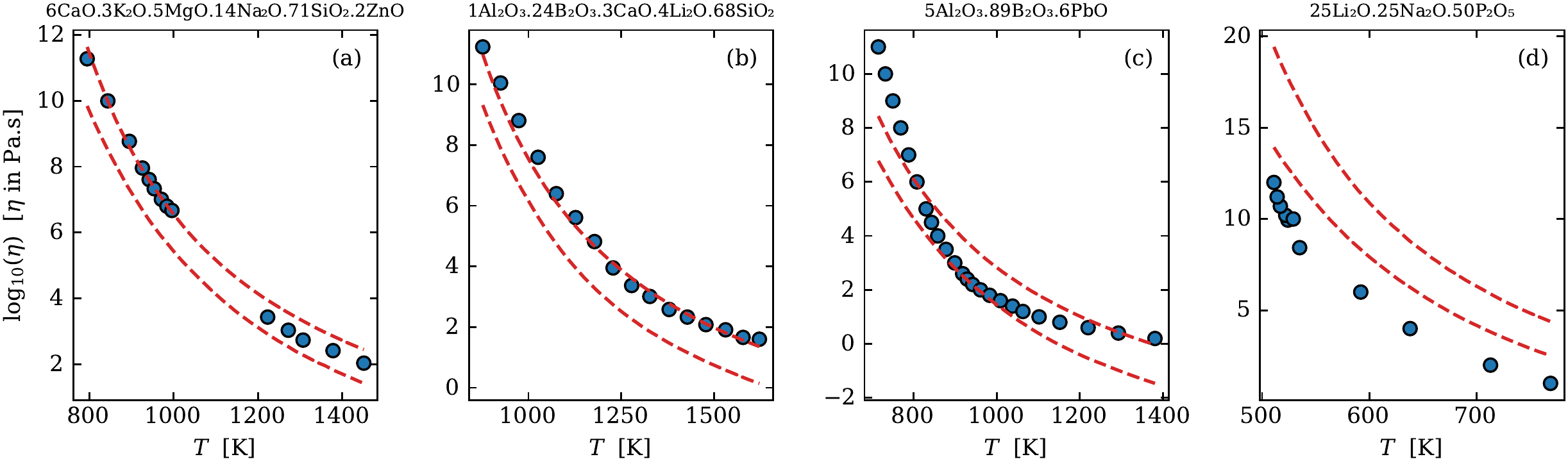}
\par\end{centering}
\caption{Base-10 logarithm of viscosity versus temperature for 4 liquids in
the test dataset. The blue circles are experimental data, and the
dashed red lines are the ViscNet prediction bands with a confidence
of \SI{95}{\percent}.}
\end{figure*}

\newpage{}

\begin{figure*}
\begin{centering}
\includegraphics[width=1\textwidth]{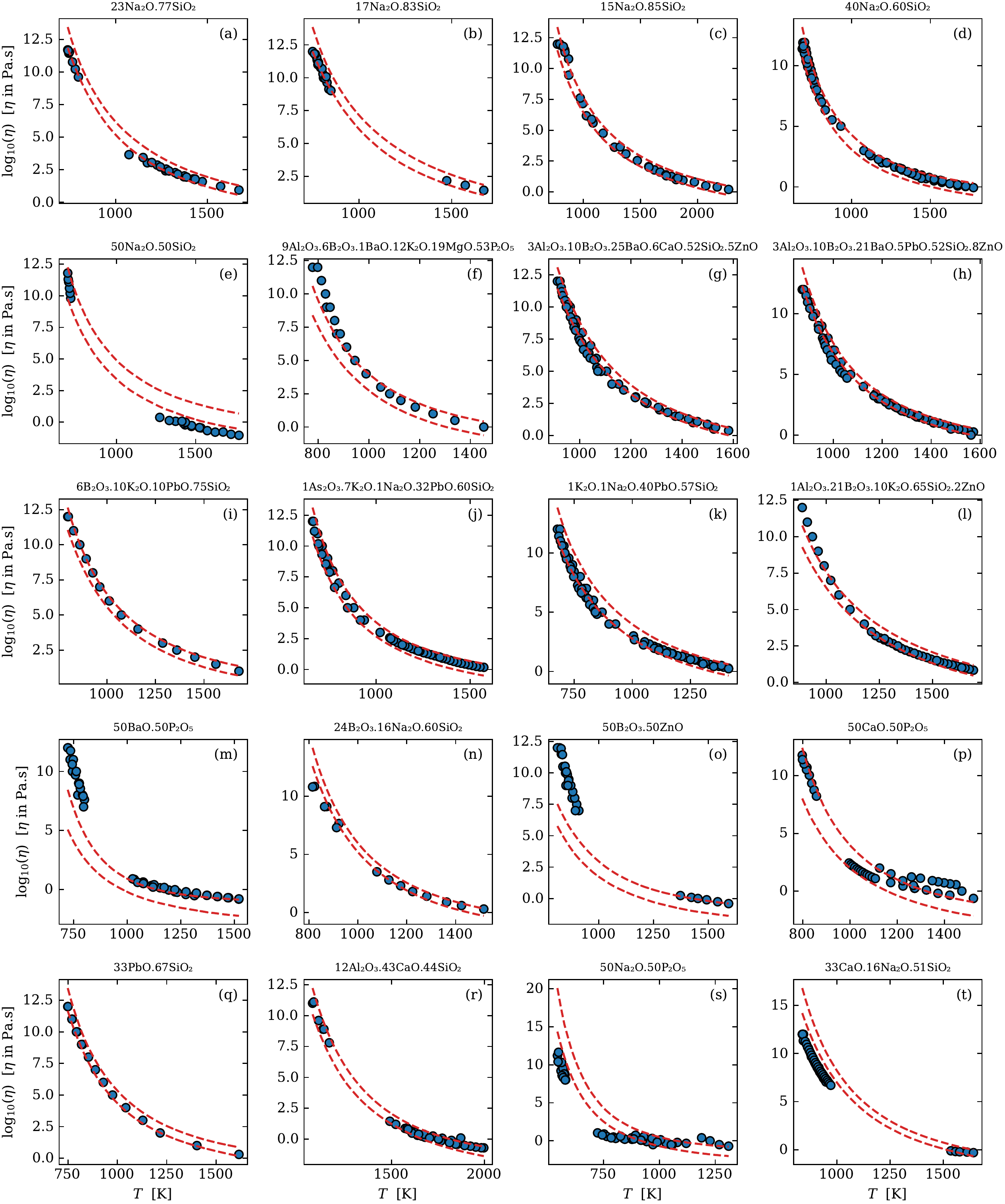}
\par\end{centering}
\caption{Base-10 logarithm of viscosity versus temperature for 20 liquids in
the test dataset. The blue circles are experimental data, and the
dashed red lines are the ViscNet-Huber prediction bands with a confidence
of \SI{95}{\percent}.}
\end{figure*}

\begin{figure*}
\begin{centering}
\includegraphics[width=1\textwidth]{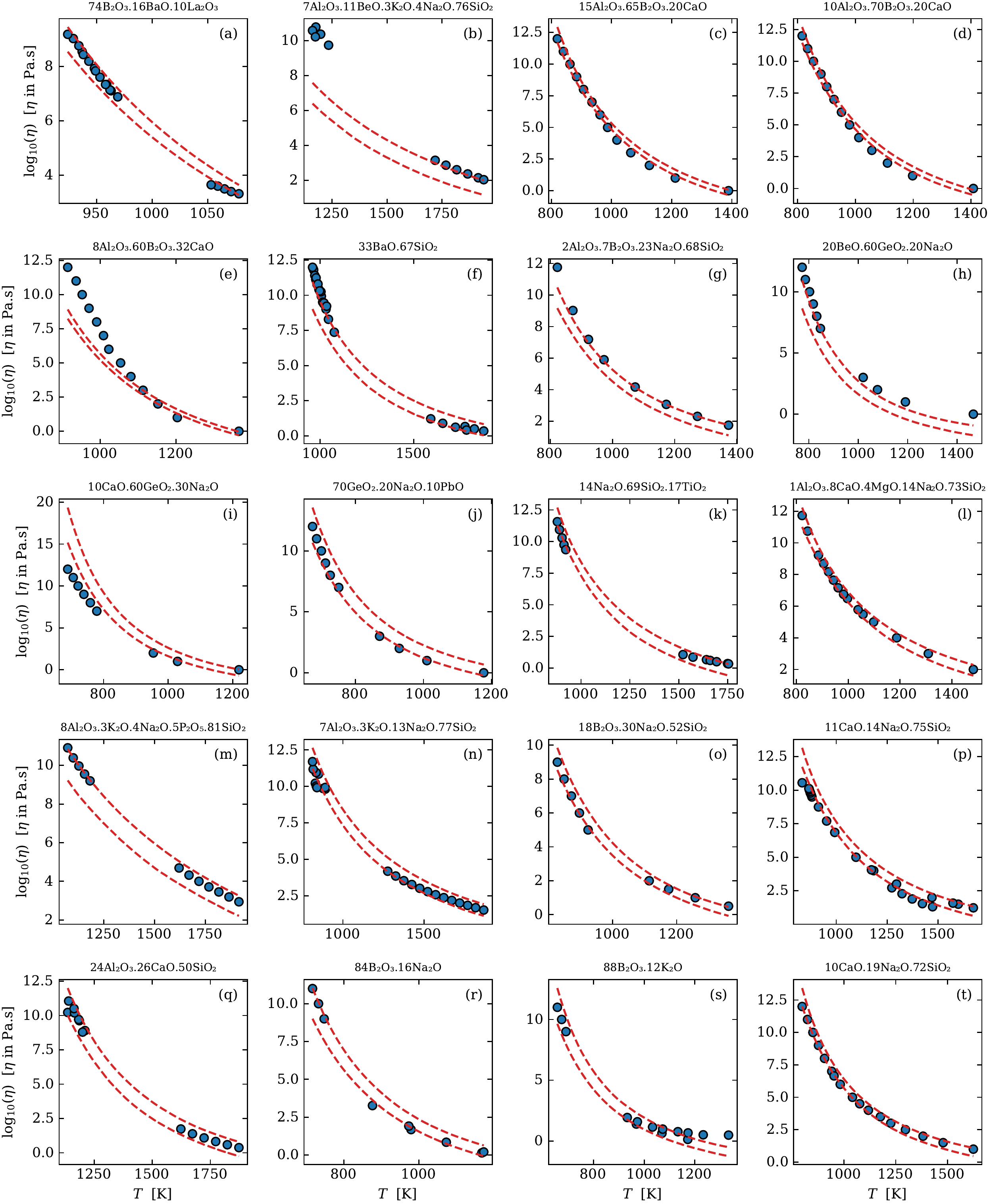}
\par\end{centering}
\caption{Base-10 logarithm of viscosity versus temperature for 20 liquids in
the test dataset. The blue circles are experimental data, and the
dashed red lines are the ViscNet-Huber prediction bands with a confidence
of \SI{95}{\percent}.}
\end{figure*}

\begin{figure*}
\begin{centering}
\includegraphics[width=1\textwidth]{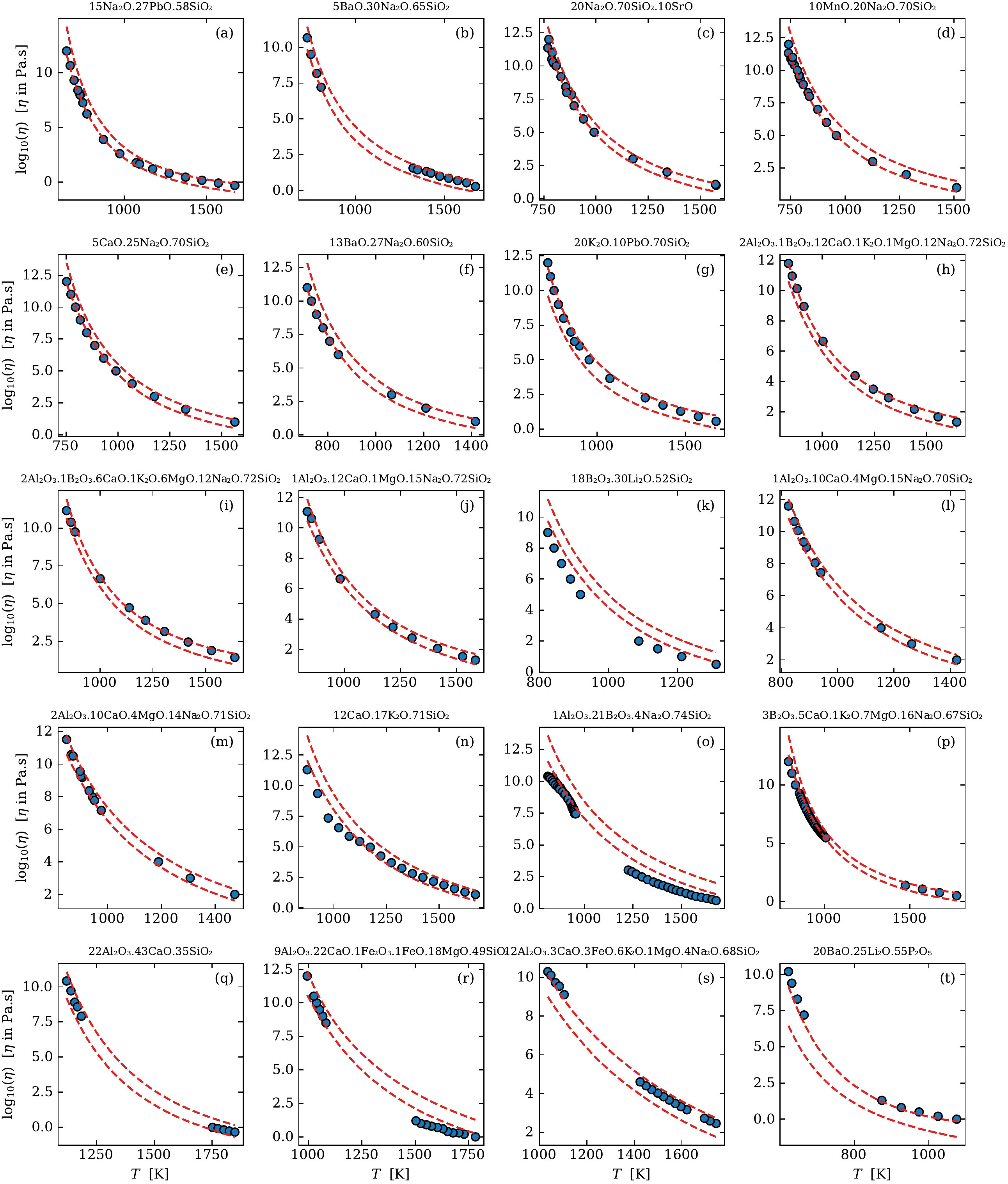}
\par\end{centering}
\caption{Base-10 logarithm of viscosity versus temperature for 20 liquids in
the test dataset. The blue circles are experimental data, and the
dashed red lines are the ViscNet-Huber prediction bands with a confidence
of \SI{95}{\percent}.}
\end{figure*}

\begin{figure*}
\begin{centering}
\includegraphics[width=1\textwidth]{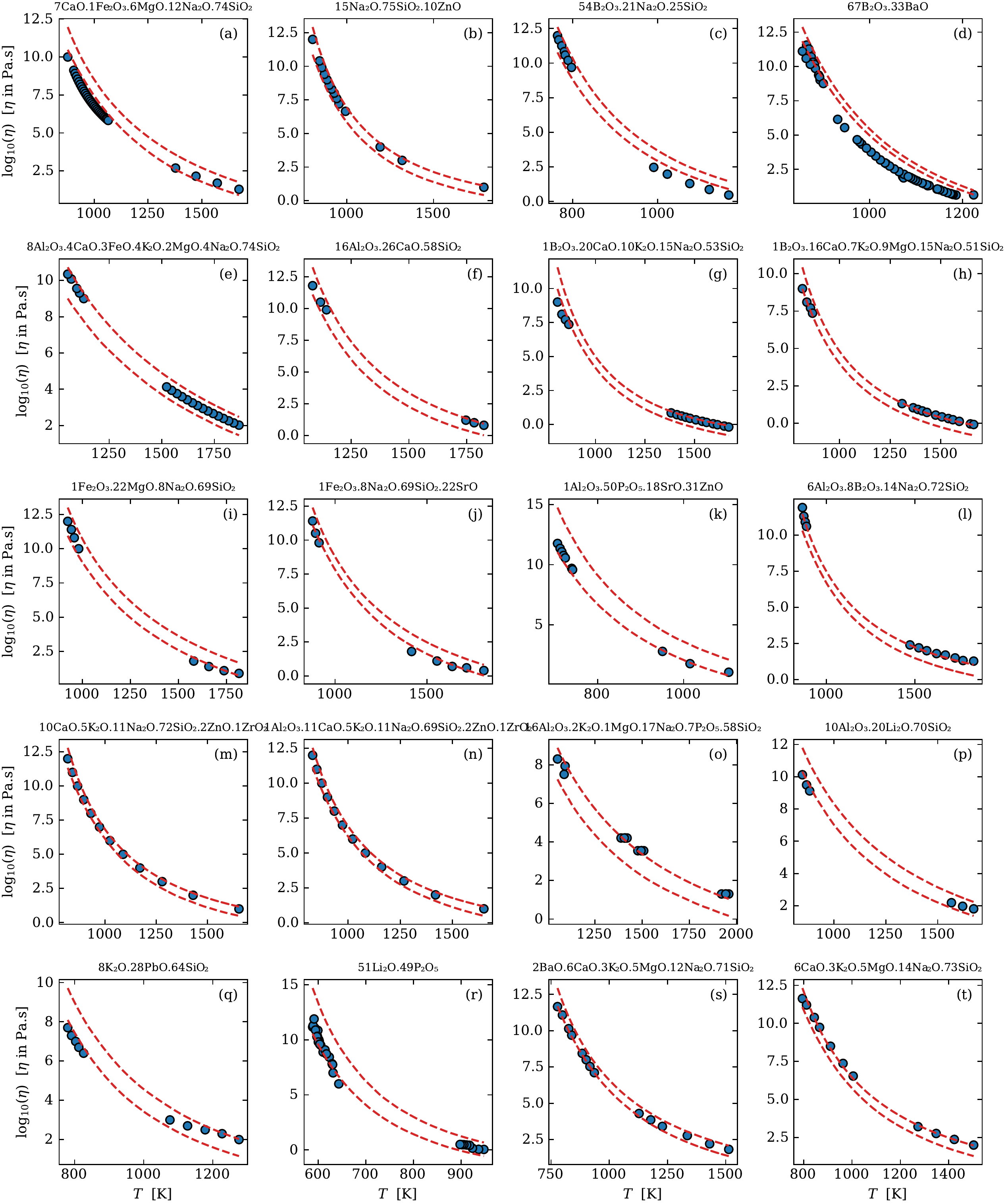}
\par\end{centering}
\caption{Base-10 logarithm of viscosity versus temperature for 20 liquids in
the test dataset. The blue circles are experimental data, and the
dashed red lines are the ViscNet-Huber prediction bands with a confidence
of \SI{95}{\percent}.}
\end{figure*}

\begin{figure*}
\begin{centering}
\includegraphics[width=1\textwidth]{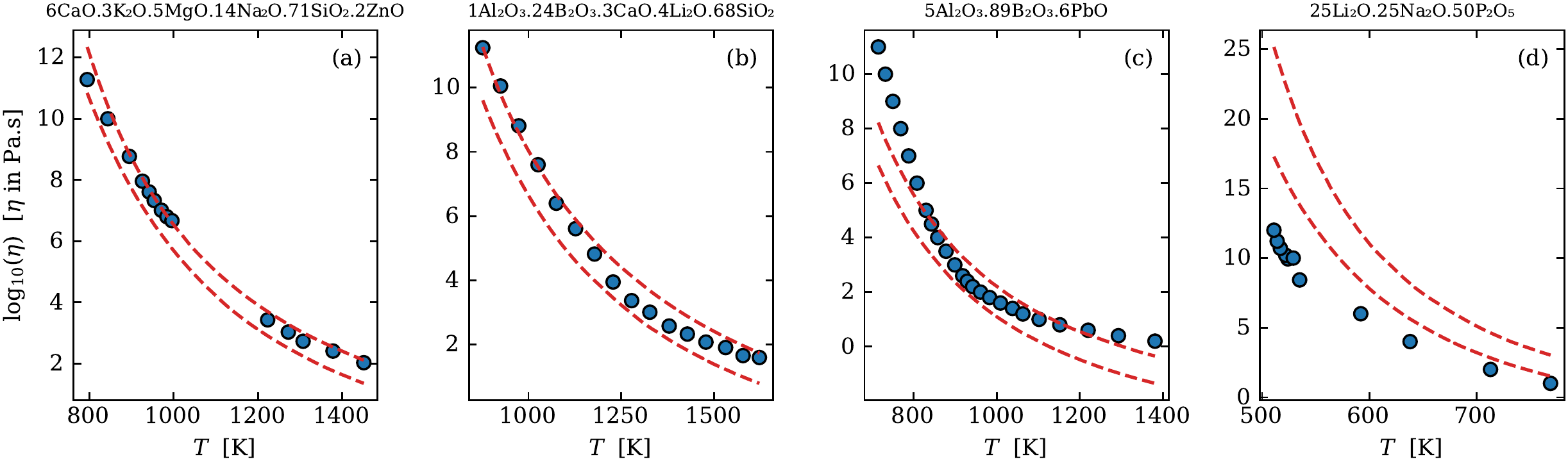}
\par\end{centering}
\caption{Base-10 logarithm of viscosity versus temperature for 4 liquids in
the test dataset. The blue circles are experimental data, and the
dashed red lines are the ViscNet-Huber prediction bands with a confidence
of \SI{95}{\percent}.}
\end{figure*}

\newpage{}

\begin{figure*}
\begin{centering}
\subfloat[]{\begin{centering}
\includegraphics[width=0.45\textwidth]{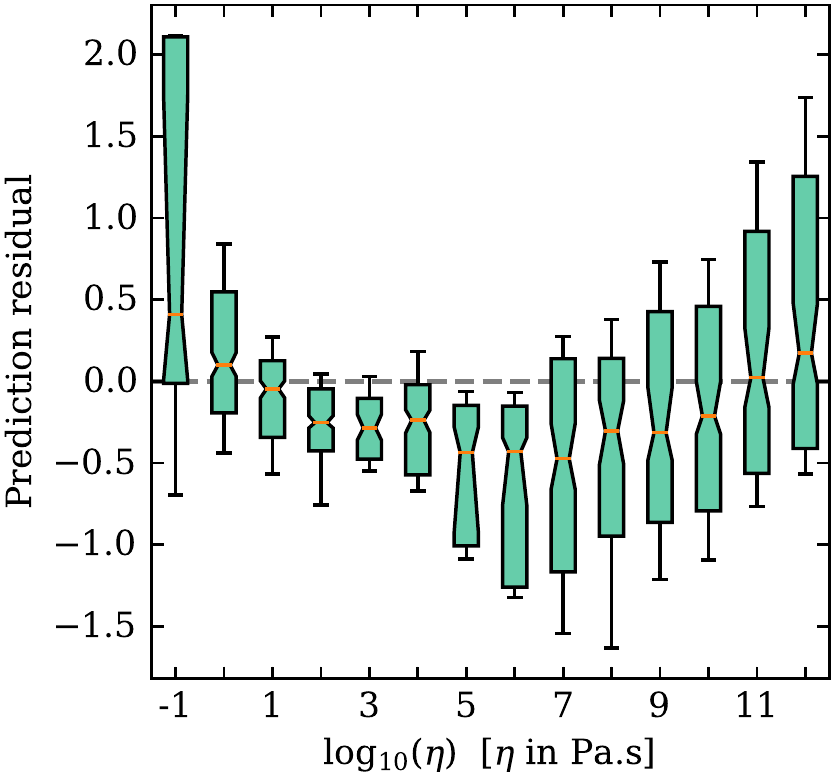}
\par\end{centering}
}\subfloat[]{\begin{centering}
\includegraphics[width=0.45\textwidth]{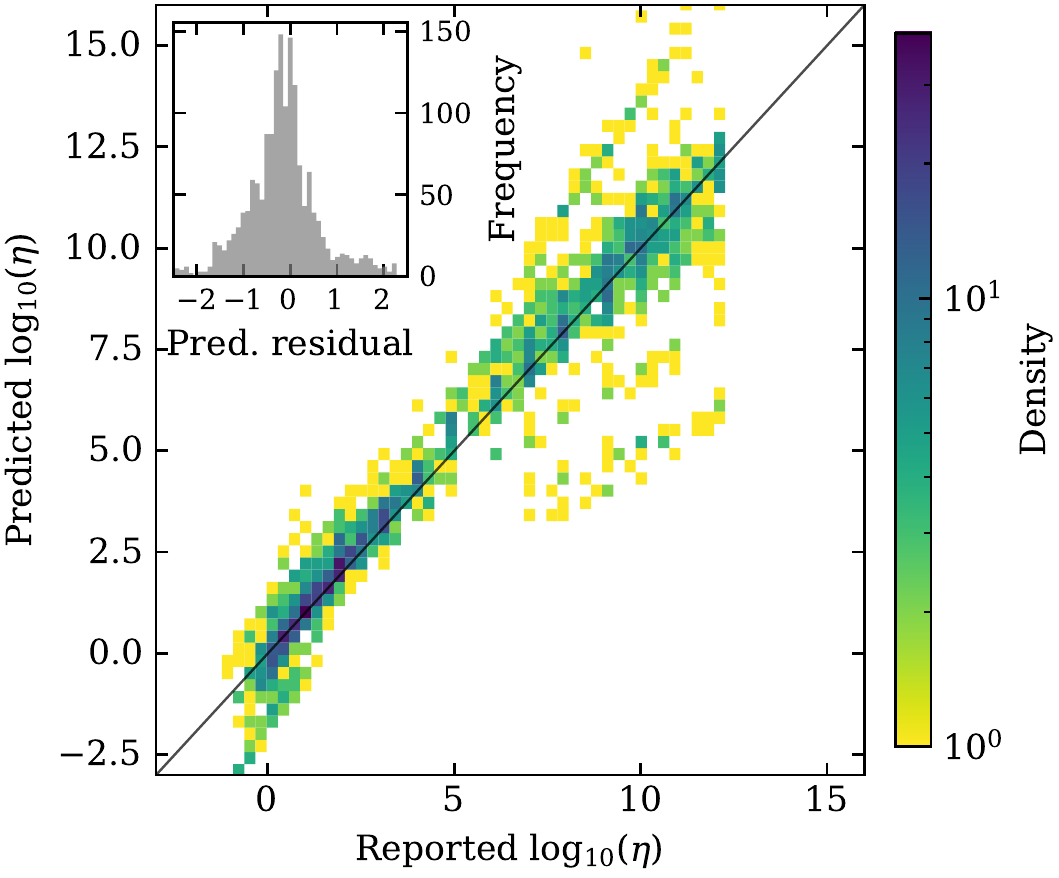}
\par\end{centering}
}
\par\end{centering}
\begin{centering}
\subfloat[]{\begin{centering}
\includegraphics[width=0.7\textwidth]{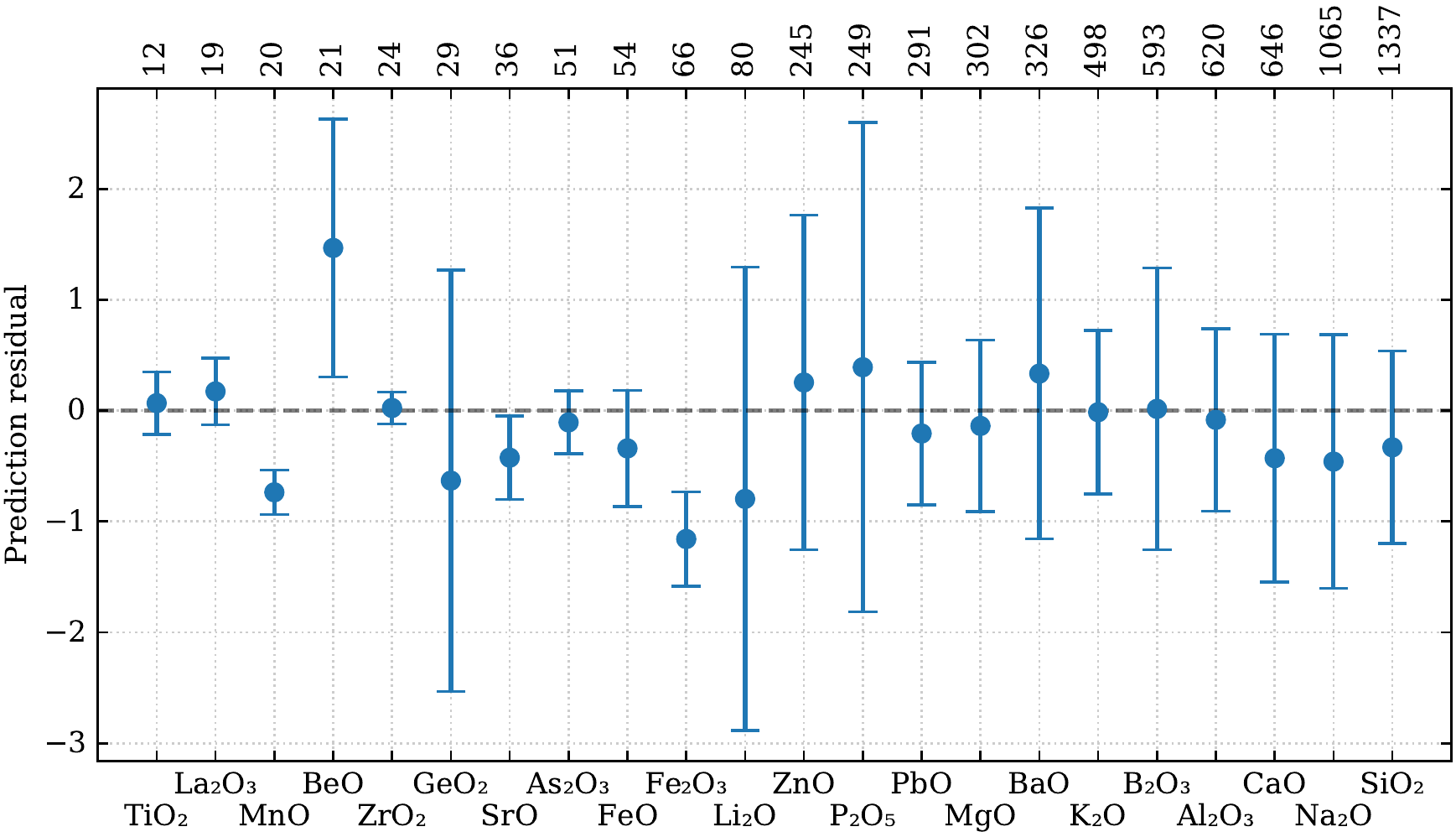}
\par\end{centering}
}
\par\end{centering}
\caption{Results of the ViscNet-VFT model for the test dataset: (\textbf{a})
boxplot of the prediction residual versus the reported value of viscosity;
(\textbf{b}) 2D histogram of predicted versus reported values of $\log_{10}\left(\eta\right)$;
and (\textbf{c}) mean and standard deviation of the prediction residual
versus the chemical compound.}
\end{figure*}

\begin{figure*}
\begin{centering}
\includegraphics[width=1\textwidth]{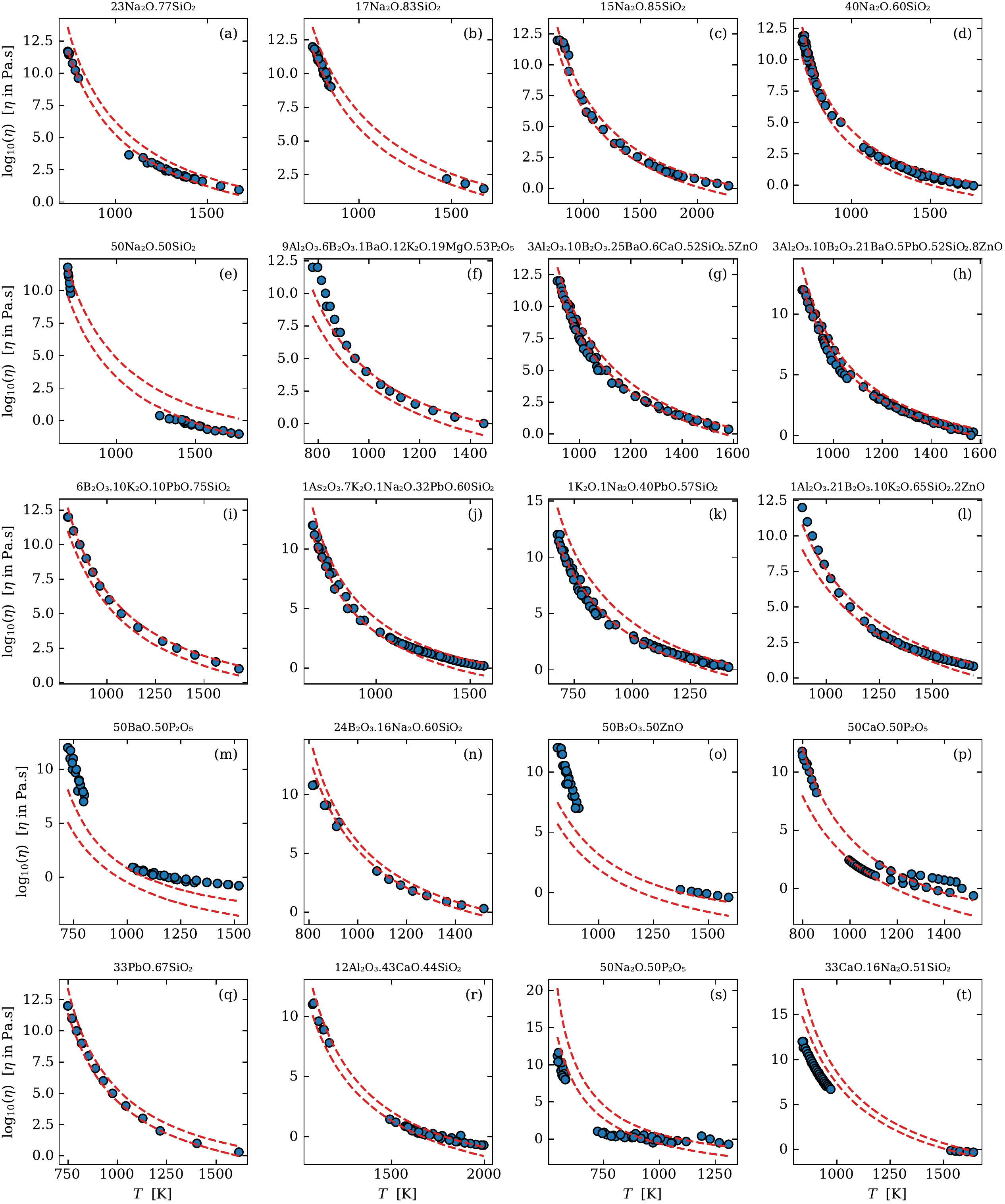}
\par\end{centering}
\caption{Base-10 logarithm of viscosity versus temperature for 20 liquids in
the test dataset. The blue circles are experimental data, and the
dashed red lines are the ViscNet-VFT prediction bands with a confidence
of \SI{95}{\percent}.}
\end{figure*}

\begin{figure*}
\begin{centering}
\includegraphics[width=1\textwidth]{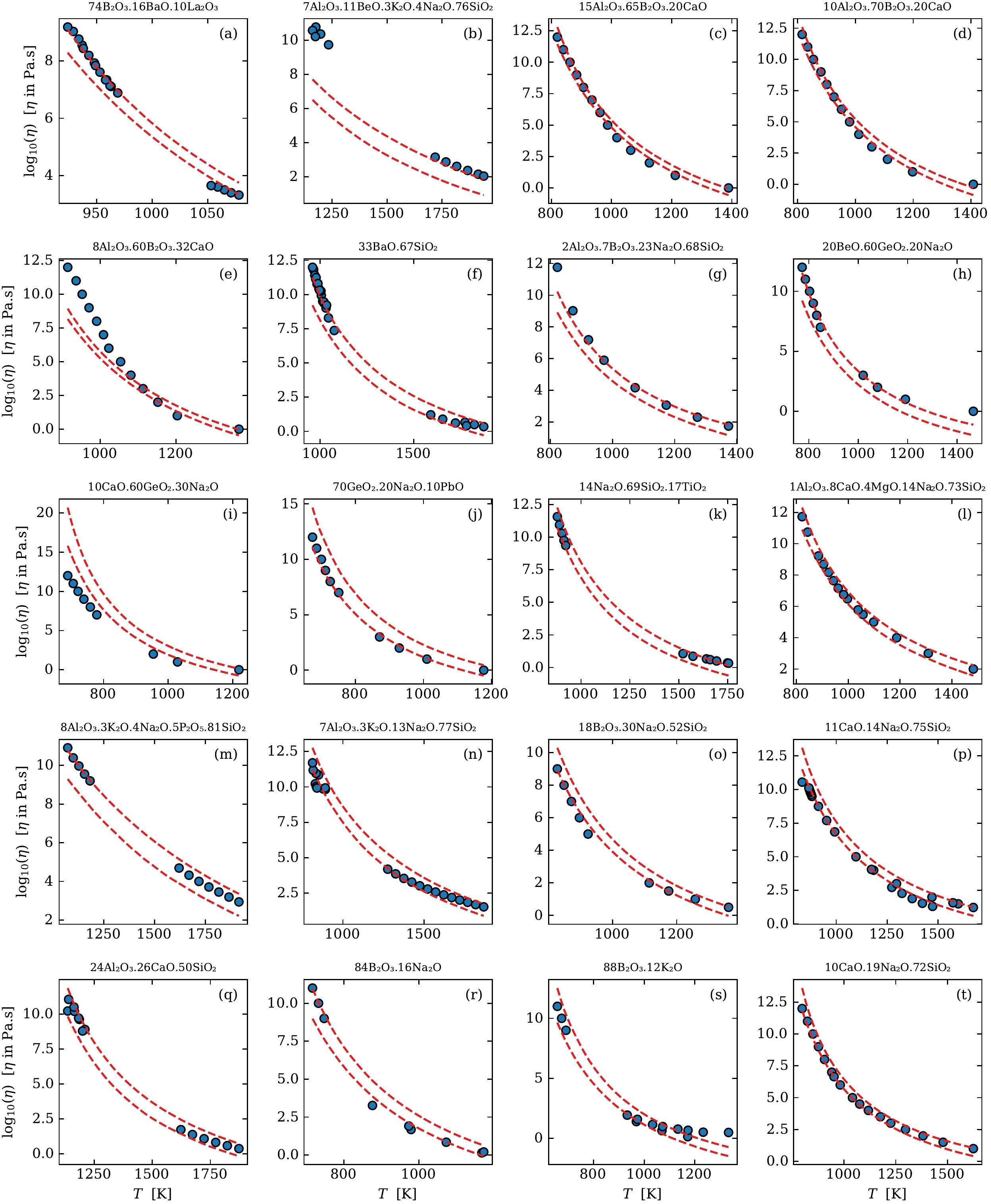}
\par\end{centering}
\caption{Base-10 logarithm of viscosity versus temperature for 20 liquids in
the test dataset. The blue circles are experimental data, and the
dashed red lines are the ViscNet-VFT prediction bands with a confidence
of \SI{95}{\percent}.}
\end{figure*}

\begin{figure*}
\begin{centering}
\includegraphics[width=1\textwidth]{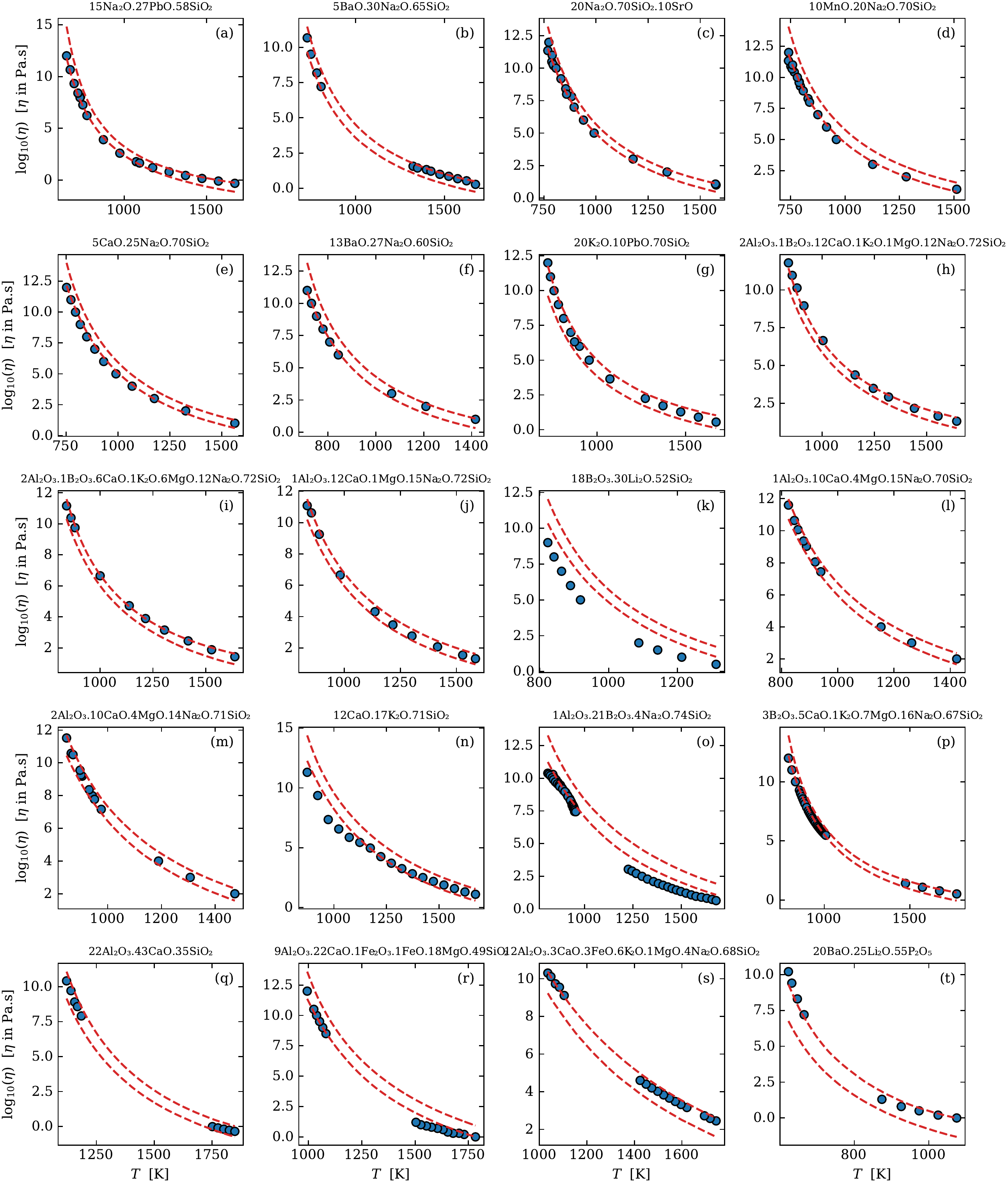}
\par\end{centering}
\caption{Base-10 logarithm of viscosity versus temperature for 20 liquids in
the test dataset. The blue circles are experimental data, and the
dashed red lines are the ViscNet-VFT prediction bands with a confidence
of \SI{95}{\percent}.}
\end{figure*}

\begin{figure*}
\begin{centering}
\includegraphics[width=1\textwidth]{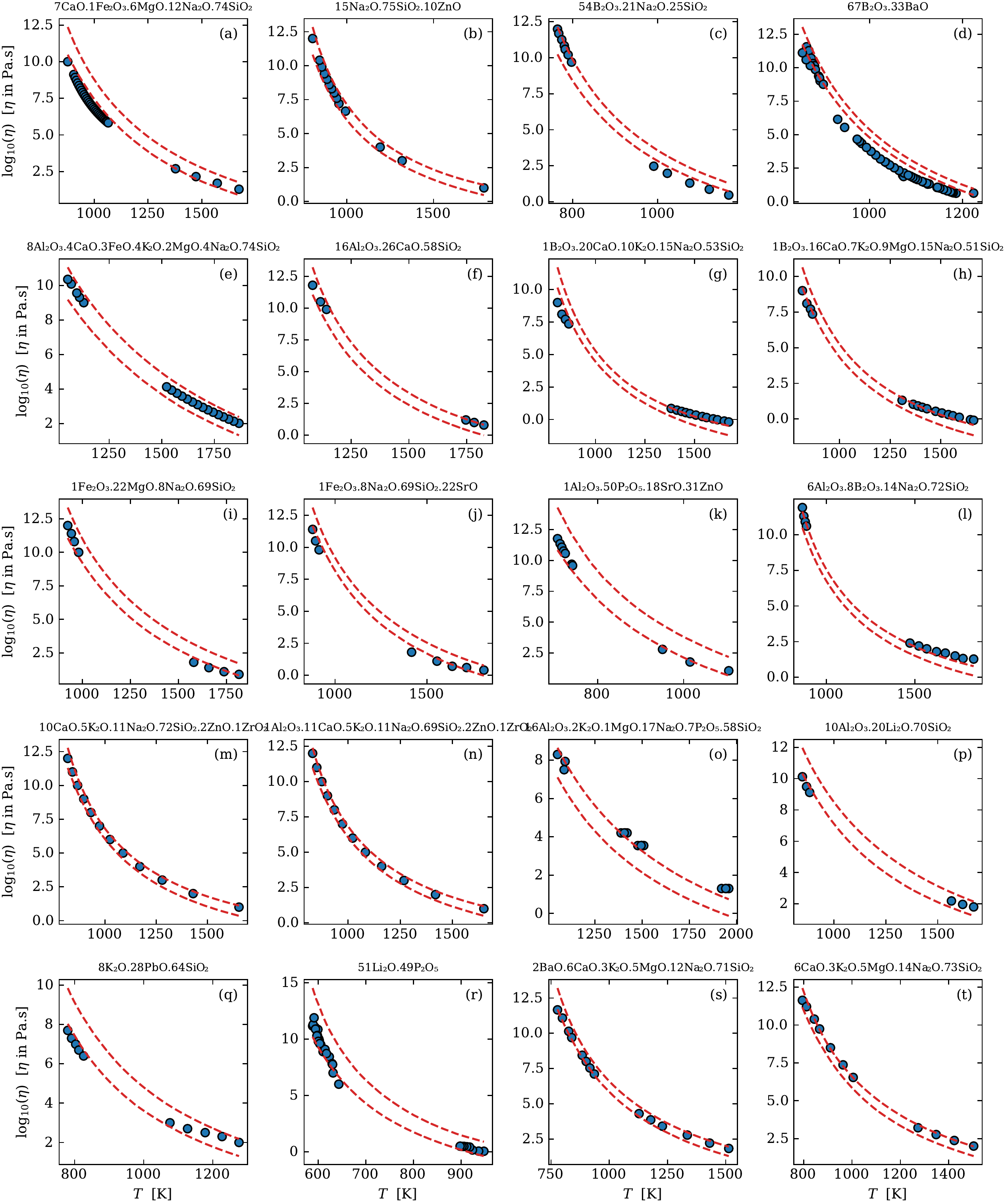}
\par\end{centering}
\caption{Base-10 logarithm of viscosity versus temperature for 20 liquids in
the test dataset. The blue circles are experimental data, and the
dashed red lines are the ViscNet-VFT prediction bands with a confidence
of \SI{95}{\percent}.}
\end{figure*}

\end{document}